\documentstyle[12pt,aaspp4,flushrt]{article}

\def\fig #1, #2, #3 {
  \smallskip
  \centerline{\psfig{figure=#1,height=#2 in,width=#3 in}} }
\def\capt{\small \baselineskip 12pt }

\def\eq#1{(\ref{eq:#1})}
\def\equ#1{equation~(\ref{eq:#1})}

\def\cl{\centerline}
\def\\{\hfill\break}

\def\vev#1{\langle#1\rangle}

\def\la{\langle}
\def\ra{\rangle}
\def\pot{{POTENT}}
\def\iras{{\it IRAS\/}}

\def\ssize{\tiny}
\def\betai{\beta_{\ssize I}}
\def\bi{b_{\ssize I}}
\def\deli{\delta_{\ssize I}}
\def\delp{\delta_{\ssize P}}
\def\delg{\delta_{g}}
\def\si{\sigma_{\ssize I}}
\def\sp{\sigma_{\ssize P}}
\def\se{\sigma_{e}}

\def\etal{{\it et al.\ }}
\def\rms{{\it rms\ }}
\def\cf{{\it cf.}}
\def\eg{{\it e.g.}}
\def\ie{{\it i.e.}}

\def\ifm#1{\relax\ifmmode#1\else$\mathsurround=0pt #1$\fi}
\def\kms{\ifmmode\,{\rm km}\,{\rm s}^{-1}\else km$\,$s$^{-1}$\fi} 
\def\hmpc{\,\ifm{h^{-1}}{\rm Mpc}}

\def\pa {\partial}

\def\ltsima{$\; \buildrel < \over \sim \;$}
\def\lsim{\lower.5ex\hbox{\ltsima}}
\def\gtsima{$\; \buildrel > \over \sim \;$}
\def\gsim{\lower.5ex\hbox{\gtsima}}
 
\def\pmb#1{\setbox0=\hbox{#1}%
 \kern-.025em\copy0\kern-\wd0
 \kern.05em\copy0\kern-\wd0
 \kern-.025em\raise.0433em\box0}
\def\vv{\pmb{$v$}}
\def\vx{\pmb{$x$}}
\def\vd{\pmb{$d$}}
\def\vr{\pmb{$r$}}
\def\vk{\pmb{$k$}}
\def\vs{\pmb{$s$}}
\def\vB{\pmb{$B$}}
\def\vL{\pmb{$L$}}
\def\v0{\pmb{$0$}}
\def\vnabla{\pmb{$\nabla$}}
\def\div{\vnabla\!\cdot\!}
\def\rot{\vnabla\!\times\!}
\def\divv{\div\vv}
\def\rotv{\rot\vv}

\begin{document}
\medskip
\baselineskip 14pt 

\title{IRAS VERSUS POTENT DENSITY FIELDS ON\\
 LARGE SCALES:~ BIASING AND OMEGA}
 
\author{Yair Sigad \altaffilmark{1}, Amiram Eldar \altaffilmark{1}, 
Avishai Dekel \altaffilmark{1,2,3},\\ 
Michael A. Strauss \altaffilmark{4,5}, Amos Yahil \altaffilmark{6}} 
\authoremail{(sigad,eldar,dekel)@astro.huji.ac.il,
strauss@astro.princeton.edu, ayahil@sbast3.ess.sunysb.edu}

\altaffiltext{1}{Racah Institute of Physics, The  Hebrew University,
Jerusalem 91904, Israel}

\altaffiltext{2}{Center for Particle Astrophysics, University of California,
Berkeley, CA 94720}

\altaffiltext{3}{Physics/Lick Observatory, University of California,
Santa Cruz, CA 95064}

\altaffiltext{4}{Princeton University Observatory,  Princeton University,
Princeton, NJ 08544}

\altaffiltext{5}{Alfred P. Sloan Foundation Fellow}

\altaffiltext{6}{Dept.\ of Physics \& Astronomy, State University of New
York, Stony Brook, NY 11794-3800}

\begin{abstract}

The galaxy density field as extracted from the \iras~1.2~Jy redshift survey
is compared to the mass density field as reconstructed by the POTENT method
from the Mark III catalog of peculiar velocities.
The reconstruction is done with Gaussian smoothing of radius $12\hmpc$, 
and the comparison is carried out within volumes of effective radii 
$31-46\hmpc$, containing $\approx\!10-26$ independent samples. 
Random and systematic errors 
are estimated from multiple realizations of mock catalogs drawn from 
a simulation that mimics the observed density field in the local universe.  
The relationship between the two density fields is found to be consistent 
with gravitational instability theory in the mildly nonlinear regime 
and a linear biasing relation between galaxies and mass.
We measure $\betai\equiv\Omega^{0.6}/\bi=0.89\pm0.12$          
within a volume of effective radius 40 \hmpc, 
where $\bi$ is the \iras\ galaxy biasing parameter at $12\hmpc$.
This result is only weakly dependent on the comparison volume, suggesting 
that cosmic scatter is no greater than $\pm 0.1$.
These data are thus consistent with $\Omega=1$ and $\bi \simeq 1$.
If $\bi > 0.75$, as theoretical models of biasing indicate, then 
$\Omega > 0.33$ at 95\% confidence.
A comparison with other estimates of $\betai$ 
suggests scale-dependence in the biasing relation for \iras\ galaxies. 

\end{abstract}

\subjectheadings{cosmology: theory --- cosmology: observation ---
dark matter --- galaxies: distances and redshifts --- 
galaxies: formation --- galaxies: clustering ---
large-scale structure of universe}

\vfill\eject

\section{INTRODUCTION}
\label{sec:intro}

A comparison of the galaxy density field derived from a redshift survey 
with the mass-density fluctuation field inferred from galaxy peculiar velocity 
data, allows one to test both gravitational instability theory (GI) 
and models for the relation between galaxies and mass.  
If the data are consistent with the assumed model, 
one can then estimate the value of the cosmological density parameter $\Omega$.
This can be illustrated, for simplicity, by the linear approximation to GI, 
for which the relation between the mass density fluctuation field 
$\delta(\vx) \equiv[\rho({\vx})-\bar\rho]/\bar\rho$ 
and the peculiar velocity field $\vv (\vx)$ is
\begin{equation}
\divv = -f(\Omega)\, \delta \ , 
\quad f(\Omega )\approx \Omega^{0.6} \ ,
\quad \vert \delta \vert \ll 1 \ ,
\label{eq:del=divv}
\end{equation}
with distances measured in \kms\ (\ie, the Hubble constant is set to unity). 
Observations of galaxy peculiar velocities allow us to measure the 
quantity on the left-hand side, while galaxy redshift surveys provide a 
measure of the {\it galaxy\/} density fluctuation field, $\delg(\vx)$.
The latter need not be identical to $\delta (\vx)$ 
(\cf, Bardeen \etal 1986; Dekel \& Rees 1987);
we adopt here the simplest toy 
model relating the two fields, linear biasing,
\begin{equation}
\delg({\vx})=b\, \delta({\vx})\ ,
\label{eq:lin_bias}
\end{equation}
where $b$ refers to the galaxies in a specific redshift survey
and at a fixed smoothing length. 
With this model for biasing, \equ{del=divv} can be rewritten 
as a relation between the observable quantities,
\begin{equation}
{\bf \nabla }\cdot {\vv}= -\beta\, \delg \ ,
\quad \beta  \equiv f(\Omega)/ b  \ .
\label{eq:gi+b}
\end{equation}
Thus, in the context of linear GI and linear biasing,
the comparison of peculiar velocities and the galaxy distribution
enables a measurement of $\beta$.                                
However, $\beta$ provides only an indirect estimate of $\Omega$,
because it is a degenerate combination of $\Omega$ and $b$.    

There is quite an extensive literature on the comparison of peculiar
velocity and redshift survey data to measure $\beta$ 
(for reviews see Dekel 1994; Strauss \& Willick 1995; Strauss 1997b; 
Dekel 1997ab; Dekel, Burstein \& White 1997   
for a general review of $\Omega$ measurements).
The comparison is not straightforward: 
the peculiar velocity data are sparse, inhomogeneously distributed, 
limited to the radial component, and quite noisy, while the 
redshift data need to be corrected to real space
and may trace the mass distribution in a 
non-trivial way. These difficulties give rise to a variety of 
statistical biases which depend on the details of the analysis carried out.  

There are two approaches to this problem, depending on whether the 
quantities that are actually compared are velocities or densities. 
Integrating both sides of \equ{gi+b} yields a predicted 
velocity field given measurements of the galaxy density field 
(equation~\ref{eq:v=int_over_x_lin} below).  
Comparison of these predictions to observed radial 
peculiar velocities allows a determination of $\beta$
(Strauss 1989; Kaiser \etal 1991; Hudson 1994).
One can make the comparison more sophisticated by smoothing the two 
velocity fields before comparing them (Davis, Nusser \& Willick 1997), 
or by  using the predicted velocity field to minimize the scatter
in the distance indicator relation from which the peculiar velocities
are measured in the first place (Strauss 1989; Roth 1994; Schlegel
1995; Shaya, Peebles, \& Tully 1995; Willick \etal 1997b).  
For \iras\ galaxies, these velocity comparisons yield 
values of $\betai$ ranging from $0.49 \pm 0.07$ (Willick \etal 1997b) 
to $0.86 \pm 0.14$ (Kaiser \etal  1991), depending on the details of 
the analysis, the smoothing scale, and the data used.
Davis \etal (1997) have claimed inconsistencies between
the peculiar velocity and redshift survey data in the context of GI
and linear biasing.

Alternatively, one can 
use the POTENT method (\S~\ref{sec:potent}) to 
recover the density fluctuation field from the peculiar velocity 
data and use \equ{gi+b}, or its nonlinear extension 
(see the discussion in \S~\ref{sec:potent_veldel}), 
to compare to the galaxy density field. 
Dekel \etal  (1993, hereafter PI93) carried out such an analysis, 
using the \iras\ 1.936 Jy redshift survey (Strauss \etal 1992b), 
and the Mark II compilation of peculiar velocities (Burstein 1989).  
The advantages of the differential form
include the facts that the direct comparison of densities is {\it local\/} 
(whereas the velocity field is sensitive to the mass distribution 
in a large volume, perhaps even 
outside the sampled volume), it is independent of reference frame, and
it allows
direct control over the smoothing of the fields.  
Monte-Carlo tests showed that the POTENT density field of PI93 
was biased in a variety of ways, forcing the use of 
an elaborate maximum likelihood technique to quantify the 
consistency between data and model, and to measure $\betai$.  
PI93 did find consistency, and concluded that 
$\betai=1.28^{+0.75}_{-0.59}$ at 95\% confidence.  
A similar comparison of POTENT densities 
with optically selected galaxies by Hudson \etal (1995) yielded  
an acceptable fit, with $\beta_{opt} = 0.74 \pm 0.13$ (1$\sigma$). 

The current paper, like PI93,  
follows the general approach of a {\it density} comparison,
but is a significant step forward 
due to a number of improvements in the quantity and quality of the 
data and the methods of analysis.  In particular:
\begin{itemize}
\item 
The current analysis is based on the Mark III catalog of 
peculiar velocities (Willick \etal  1995, 1996, 1997a).
With $\sim\!3400$ galaxies, it is more than three times bigger than the 
Mark II data set, and has better space coverage.
The data sets composing the Mark III catalog have been treated with more care,
especially in self-consistently calibrating the Tully-Fisher (TF) relations,
grouping, and correcting for inhomogeneous Malmquist bias 
(\S~\ref{sec:potent}). 
\item 
The POTENT method for deriving the density fluctuation field $\delp$
from peculiar velocity data has been much improved since PI93
(\S~\ref{sec:potent}; Dekel \etal 1997).  
\item 
The current analysis uses the \iras\ 1.2 Jy redshift survey 
(Fisher \etal  1995), 
containing twice as many galaxies as in the 1.936 Jy survey.
\item
The method for deriving a uniform galaxy density field $\deli$ 
from the redshift 
data has been improved in several ways since PI93 (\S~\ref{sec:iras}).
\item 
The availability of much more realistic simulations of both the
\iras\ and Mark III datasets (\S~\ref{sec:eval_mock}; Kolatt \etal 1996) 
allows much better error analysis in both the peculiar velocity and 
density fields, and in the comparison. 
\end{itemize}
We use these simulations to assess biases in our determination of $\betai$.  
Unlike PI93, these biases turn out to be negligible, allowing us to 
sidestep the rather elaborate likelihood analysis of that paper.  
Indeed, we will use the simulations themselves as a guide to whether our 
data are statistically consistent with the null hypothesis of GI and linear
biasing. 
We also use them to quantify the statistical errors in our final results. 

The current analysis compares the density fields smoothed with a 
Gaussian window of radius $12\hmpc$, where the fluctuations are 
of order unity and therefore require a mildly nonlinear treatment. 
The POTENT analysis computes the density field $\delp$,
a generalization of $-f(\Omega)^{-1} \divv$ that is a nonlinear 
function of $\Omega$ and the spatial partial derivatives of the
observed $\vv(\vx)$ (equation~\ref{eq:delc+}).
The \iras\ reconstruction, in turn, yields a mildly nonlinear galaxy density
field, $\deli$, that is a weak function of $\Omega$ and $\bi$ --- only 
via the corrections from redshift space to real space 
(equations~[\ref{eq:cz-r}] and [\ref{eq:v=int_over_x}]).
Equation \eq{gi+b} is thus replaced by
\begin{equation}
\deli = \bi\, \delp \ .    
\label{eq:deli=delp}
\end{equation}
The $\Omega$ dependence of equation~(\ref{eq:gi+b}) is already included 
in $\delp$.  
The density fields are recovered from the data for assumed values
of $\Omega$ and $\bi$. Then, $\bi$ is determined by \equ{deli=delp},
and $\betai$ is quoted.
The analysis is carried out for several initial values of $\Omega$ 
and $\bi$ to confirm the robustness of the estimate of $\betai$.

The present work is less ambitious than PI93 in one respect. 
An attempt was made in PI93 to use the nonlinear effects
to break the degeneracy between $\bi$ and $\Omega$, 
but with only limited success. The resulting constraints on each parameter 
separately were quite weak, indicating that the nonlinear effects
associated with these data are not sufficient for this purpose. 
Even with the new data, the nonlinear effects are 
comparable to the errors that accompany the reconstructions. 
Furthermore, if we are to consider nonlinear gravity, we should also allow 
nonlinear extensions to the biasing relation, \equ{lin_bias}, 
but in practice, it is not clear how to distinguish
nonlinear gravity effects from nonlinear biasing effects.
We therefore limit ourselves in this paper to determining the
degenerate combination $\betai$.
 
The outline of this paper is as follows: 
In \S~\ref{sec:potent} we discuss the POTENT reconstruction of the
mass-density field from peculiar velocities, and in \S~\ref{sec:iras} 
the reconstruction of the galaxy density field from the \iras\ redshift
survey. 
In \S~\ref{sec:eval} 
we use mock catalogs to evaluate the random and systematic errors in the 
reconstructed density fields, and to minimize them if possible. 
In \S~\ref{sec:method} we describe our method of comparison of the 
two fields, and estimate the systematic and random errors in the 
measurement of $\betai$.
In \S~\ref{sec:results} we perform the comparison of the real data, 
evaluate goodness of fit, and determine the value of $\betai$.
In \S~\ref{sec:conc} we conclude with the implications of our results
and compare them with other recent determinations of $\betai$.

\section{POTENT RECONSTRUCTION FROM PECULIAR VELOCITIES}
\label{sec:potent}

The POTENT procedure recovers the underlying mass-density fluctuation field 
from a whole-sky sample of observed radial peculiar velocities.
The steps involved are: 
(a) preparing the data for POTENT analysis, including grouping and 
    correcting for Malmquist bias,
(b) smoothing the peculiar velocities into a uniformly-smoothed radial 
    velocity field with minimum bias,
(c) applying the ansatz of gravitating potential flow to recover
    the potential and three-dimensional velocity field, and
(d) deriving the underlying density field $\delp$ by an approximation to GI
    in the mildly nonlinear regime.
The revised POTENT method, which grew out of the original method of 
Dekel, Bertschinger \& Faber (1990, DBF), is described in detail in Dekel 
\etal (1997) and is reviewed in the context of other methods by Dekel 
(1997a, 1997b).  
We emphasize below the important improvements since PI93.  

\subsection {Mark III Data}
\label{sec:potent_m3}

We use the most comprehensive catalog of peculiar velocity data available 
today, the Mark III catalog
(Willick \etal 1995, 1996, 1997a), which is a careful
compilation of several data sets of spiral and elliptical galaxies
under the assumption that all galaxies
trace the same underlying velocity field. 
The non-trivial procedure of merging the data sets 
accounts for differences in the selection
criteria, the quantities measured, the method of measurement and
the TF calibration techniques.
The data per galaxy consist of a redshift $z$, a
``forward" TF (or $D_n-\sigma$) inferred distance, $d$, and an error. 
The radial peculiar velocity is then $u=cz-d$.

The Mark II data used in PI93 consisted of about 1000 galaxies
(mostly ellipticals by  Lynden-Bell \etal 1988 and spirals by
Aaronson \etal 1982).
The extended Mark III catalog consists of $\sim\!3400$ galaxies,
including the Mark II data (with improvements to the uniformity of 
Aaronson \etal 1982 by Tormen \& Burstein 1995) 
as well as newer spiral data sets by Courteau (1992), Willick (1991),
Mathewson, Ford, \& Buchhorn (1992), and Han \& Mould (1992, and
references therein).
This sample enables a reasonable recovery of the dynamical fields with
$12\hmpc$ smoothing in a sphere of radius $\sim\!60\hmpc$ about
the Local Group (LG).

\subsection {Correcting Malmquist Bias}
\label{sec:potent_im}

Even after the selection bias in the TF calibration is properly corrected,
the inferred distance $d$ for a given galaxy, and therefore its peculiar 
velocity, suffer from a Malmquist bias due to the cross-talk between the
random distance errors and the distribution of galaxies about $d$.
We correct this bias in a statistical way before using the velocities
as input to POTENT.
If the scatter around the TF relation is Gaussian (Willick \etal 1997a), 
i.e., the absolute magnitude $M$ is distributed normally for a given 
log-linewidth $\eta$ with standard deviation $\sigma_m$, 
then the forward inferred distance of a galaxy at a true distance $r$
is distributed log-normally about $r$, with relative error
$\Delta\!\approx\! 0.46\sigma_m$.
Given $d$, the expectation value of $r$ is 
\begin{equation}
E(r \vert d)=
{ \int_0^\infty r P(r\vert d) dr
\over
\int_0^\infty P(r\vert d) dr } =
{ \int_0^\infty r^3 n(r)\ {\rm exp}
\left( -{[{\rm ln}(r/d)]^2 \over 2\Delta^2} \right) dr
\over
\int_0^\infty r^2 n(r)\ {\rm exp}
\left( -{[{\rm ln}(r/d)]^2 \over 2\Delta^2} \right) dr } \ ,
\label{eq:malmquist}
\end{equation}
where $n(r)$ is the number density in the underlying distribution from
which the galaxies were selected.
The deviation of $E(r\vert d)$ from $d$ is the bias.
The homogeneous part ($n\!=\!$ constant), which arises from the 
three-dimensionality of space, 
reduces to $E(r\vert d) = d\, e^{3.5 \Delta^2}$,
and thus is easy to correct.  
 
Fluctuations in $n(r)$ are responsible for the inhomogeneous Malmquist bias
(IM), which, if uncorrected, will 
systematically enhance the inferred density perturbations $\delp$,
and thus the deduced value of $\beta$.
The galaxies are first grouped in $z$-space (Willick \etal 1996), reducing 
the distance error of each group and thus weakening the bias.
The inferred distance of each object is then replaced by 
$E(r\vert d)$, with an assumed $n(r)$ properly corrected for grouping and 
for redshift cutoffs in the data.
We use high-resolution density fields of \iras\ and optical galaxies 
(Hudson 1993) for the required input $n(r)$ for spirals and 
ellipticals respectively. 
Tests with mock data demonstrated that 
the IM bias can be reduced to the level of a few percent in $\delp$ 
(Dekel \etal 1997; Eldar, Dekel \& Willick 1997).

The resultant correction to $\delp$ is less than $20\%$ even at the highest 
peaks, and is much smaller throughout most of space.
Thus, the use of \iras\ data for the IM correction of the 
Mark III data introduces only a weak coupling to the POTENT output 
density field. For the purpose of this paper, the POTENT and \iras\ 
density fields can thus be regarded as essentially independent data sets.
Indeed, a different IM correction procedure in the framework of an 
``inverse" TF analysis, which does not rely on external input for $n(r)$, 
yields a density field that is consistent with the result of the forward 
analysis applied here (Dekel \etal 1997; Eldar, Dekel \& Willick 1997; 
following Landy \& Szalay 1992).

\subsection{Smoothing the Radial Velocities}
\label{sec:potent_twf}

The goal of the POTENT analysis is to use the set of 
radial peculiar velocities $u_i$
at positions $\vd_i$ to determine the underlying continuous  
fields $\vv(\vx)$ and $\delta(\vx)$,
both smoothed with a Gaussian of radius $R_s$. 
We denote hereafter
a 3D Gaussian smoothing window of radius $R$ by G$R$, such that $12\hmpc$ 
smoothing is indicated by G12, and so on.  
The first, most difficult step is the smoothing, or interpolation, 
into a continuous radial velocity field $u(\vx)$. 
The goal is minimum bias compared to
the radial component of the true field, had it been sampled
noiselessly, densely and uniformly, and perfectly smoothed in 3D. 
The radial velocity field at $\vx_c$ is taken to be the value 
of an appropriate {\it local} velocity model $\vv(\alpha_k,\vx\!-\!\vx_c)$ 
at $\vx\!=\!\vx_c$. The model parameters $\alpha_k$ are
obtained by minimizing the weighted sum of residuals,
\begin{equation}
\sum_i W_i\,
[u_i-\hat{\vx}_i\cdot\vv(\alpha_k,\vx _i)]^2 \ ,
\label{eq:sumv}
\end{equation}
within an appropriate local window $W_i\!\equiv\!W(\vx_i,\vx_c)$.
The window is a Gaussian, modified such that it
minimizes the combined effect of the following three types of errors.

\noindent{\it Tensor window bias.}
Because the radial peculiar velocity vectors are not parallel, 
the $u_i$'s cannot be averaged as scalars,
and $u(\vx_c)$ requires a fit of a local 3D velocity model.
The POTENT version of DBF used the simplest local model with three parameters, 
\begin{equation} 
\vv(\vx)=\vB \ ,
\label{eq:vel-local-bulk} 
\end{equation}
for which the solution can be expressed explicitly in terms of a 
tensor window function.
However, a bias occurs because
the tensorial correction to the spherical window has conical symmetry,
and because variations of $\vv$ within the window, 
such as a spherical infall pattern, 
may be interpreted erroneously as a local bulk velocity. 
This bias amounts to $\sim 300 \kms$ at the Great Attractor (see below).
The window bias can be reduced by introducing shear into the model,
\begin{equation} 
\vv(\vx)=\vB + \bar{\bar{\vL}} \cdot (\vx-\vx_c) \ ,
\label{eq:vel-local-shear} 
\end{equation}
where $\bar{\bar{\vL}}$ is a symmetric tensor that automatically ensures
local irrotationality.  The additional six  terms tend to ``absorb" most of
the bias, leaving $\vv(\vx_c)\!=\!\vB$ less biased.
Unfortunately, a high-order model tends to
pick undesired small-scale noise.  The optimal compromise for the
Mark III data was found to be \equ{vel-local-shear} 
out to $r\!=\!40\hmpc$, smoothly changing to 
\equ{vel-local-bulk} beyond $60\hmpc$ (Dekel \etal 1997).

\noindent{\it Sampling-gradient bias.}
Gradients in the true velocity field within the effective window, 
coupled with the non-uniform sampling, will introduce a bias 
if the smoothing is not equal-volume weighted. 
We thus weight each object in proportion to the local volume that it 
``occupies", estimated by $V_i\!\propto\!R_{4,i}^3$, where $R_4$ 
is the distance to the 4-th nearest object.
Simulations show that this reduces the bias to negligible levels
for $r < 60\hmpc$, except for the very sparsely sampled Galactic zone 
of avoidance.
We will use the $R_4(\vx)$ field as a flag for poorly sampled regions,
to be excluded from the comparison with \iras.

\noindent{\it Random errors.}
The ideal weighting for reducing the effect of Gaussian noise
would be  $W_i\!\propto\!\sigma_i^{-2}$, where $\sigma_i$ are the 
distance errors. Unfortunately, this weighting spoils the 
volume weighting.  As a compromise we weight by both, \ie
\begin{equation}
W(\vx_i,\vx_c) \propto
V_i\, \sigma_i^{-2}\, \exp [-(\vx_i-\vx_c)^2/ 2R_s^2] \ .
\label{eq:window}
\end{equation}

\subsection{From Radial Velocity to Density Field}
\label{sec:potent_veldel}

Under GI, the large-scale velocity field 
is expected to be irrotational, $\rotv\!=\!0$.  
This remains a good approximation in the mildly nonlinear regime as
long as the field is properly smoothed
(Bertschinger \& Dekel 1989; DBF).  
This implies that the velocity field can be derived from a scalar potential,
$\vv(\vx)\!=\!-\vnabla\Phi(\vx)$, and thus
the potential can be computed by integration along the lines of sight,
\begin{equation}
\Phi(\vx) = -\int_0^r u (r',\theta,\phi) dr' \ .
\label{eq:velpot}
\end{equation}
The two missing transverse velocity components are then
recovered by differentiation.

The subsequent derivation of the underlying mass-density fluctuation field 
requires a solution to the equations of GI
in the mildly nonlinear regime.
The linear approximation is limited to the small
dynamical range between a few tens of megaparsecs
and the $\sim\!100\hmpc$ extent of the current samples.
Our current peculiar velocity samples enable us to perform reliable 
dynamical analyses with a smoothing radius as small as
$\sim\!10\hmpc$, where $\vert\divv\vert$ reaches values
larger than unity and therefore nonlinear effects play a role.
We appeal to the Zel'dovich (1970) approximation, which is known to be
a successful tool in the mildly nonlinear regime. Substituting an Eulerian 
version of the Zel'dovich approximation in the continuity equation yields 
(Nusser \etal 1991)
\begin{equation}
\delta_c(\vx) = \Vert \pmb{I} - f^{-1} {\pa \vv / \pa \vx} \Vert -1 \ ,
\label{eq:delc}
\end{equation}
where the bars denote the Jacobian determinant and $\pmb I$ is the unit
matrix. 
This was the approximation used in PI93. Tests with $N$-body simulations 
show that it does an excellent job for $\delta \geq 0$, but it tends 
to be an overestimate for $\delta < 0$ 
(Mancinelli \etal 1994; Ganon \etal 1997). 

The Zel'dovich displacement is first order in $f^{-1}$ and
$\vv$, and therefore the determinant in $\delta_c$ includes
second- and third-order terms as well,
involving sums of double and triple products of partial derivatives:
\begin{equation}
\delta_{c} = -f^{-1} \divv
             +f^{-2} \Delta_2
             +f^{-3} \Delta_3 \ ,
\label{eq:delc-expand}
\end{equation}
where 
\begin{equation}                                                   
\Delta_2(\vx)= \sum_{i < j} \left[\left({\pa v_i \over \pa x_j}\right)^2
- {\pa v_i \over \pa x_i}
{\pa v_j \over \pa x_j} \right]                                  
\label{eq:d2}                                                     
\end{equation}                                                 
and                                                           
\begin{equation}                                             
\Delta_3(\vx) = \sum_{i,j,k} \left[ 
{\pa v_i \over \pa x_i}{\pa v_j \over \pa x_k}{\pa v_k \over \pa x_j} - 
{\pa v_1 \over \pa x_i}{\pa v_2 \over \pa x_j}{\pa v_3 \over \pa x_k} \right] ,
\label{eq:d3}                                                 
\end{equation}                                               
where the sum is over the three cyclic permutations of $(i,j,k) =
(1,2,3)$. 
The approximation can be improved by slight adjustments to the coefficients
of the three terms in \equ{delc-expand},
\begin{equation}
\delta_{c+} = -(1+\epsilon_1) f^{-1} \divv
             +(1+\epsilon_2) f^{-2} \Delta_2
             +(1+\epsilon_3) f^{-3} \Delta_3 \ .
\label{eq:delc+}
\end{equation}
These coefficients were empirically tuned to best fit a family of CDM 
simulations of $12\hmpc$ smoothing over the whole range of $\delta$ values,
with $\epsilon_1=0.06$, $\epsilon_2=-0.13$ and $\epsilon_3=-0.3$.
This approximation is found to be robust to unknown quantities 
such as the value of $\Omega$, the shape of the power spectrum,
and the degree of nonlinearity as determined by the fluctuation amplitude
and the smoothing scale (Ganon \etal 1997).
We adopt $\delp=\delta_{c+}$, \equ{delc+}, in this paper. 

\section{THE \pmb{$\iras$}\ RECONSTRUCTION}
\label{sec:iras}

The redshifts of galaxies in the \iras\ sample differ from the true distances
by the same peculiar velocities that one is attempting to measure in the Mark
III dataset,  
\begin{equation} 
cz = r + \hat{\vx}\cdot\vv(\vx)\ ,
\label{eq:cz-r} 
\end{equation}
where $cz$ and $\vv$ are measured in the same frame of reference. 
Because of peculiar velocities, 
the galaxy density field measured in redshift space, $\delg(\vs)$, differs
systematically from that in real space, $\delg(\vx)$
(Kaiser 1987;  
see reviews in Dekel 1994; Strauss \& Willick 1995; Strauss 1997b). 
Gravitational instability theory
enables us to correct for the effects of these velocities.
Given the divergence field $\divv$ and appropriate boundary conditions, 
the velocity is
\begin{equation}
{\vv}({\vx})=-{1\over {4\pi }}
\int \divv({\vx'})\,{{{\vx'}-{\vx}} \over {{\left| {\vx'}-{\vx}\right|}^3}}\,
  d^3\vx'\ .
\label{eq:v=int_over_x}
\end{equation}
In the linear regime, \equ{gi+b} yields
\begin{equation} 
{\vv}({\vx})={\beta\over {4\pi }} 
\int \delg(\vx')\,{{{\vx'}-{\vx}} \over {{\left|  
{\vx'}-{\vx}\right|}^3}}\, 
  d^3\vx'\ . 
\label{eq:v=int_over_x_lin} 
\end{equation} 
Of course, the input density field is  
defined in real space, which requires that we find a simultaneous solution for
the real space density and velocity fields.
Yahil \etal (1991, hereafter YSDH) and Strauss
\etal (1992a) describe an iterative technique which implements
equations \eq{cz-r} and \eq{v=int_over_x_lin} 
to calculate the peculiar velocity field and convert the redshifts
to distances for an assumed value of $\beta$.
Our current implementation of the iteration procedure 
has three new features: the treatment of triple-valued zones, 
the filtering of the density field, and the nonlinear corrections. 

\subsection{Modeling Triple-Valued Zones}
\label{sec:iras_tvz}

  Given a model for the 
velocity field from equations~\eq{gi+b} and \eq{v=int_over_x},
one can compute the redshift-distance relation along a line of sight
to a given galaxy from \equ{cz-r}.  This can be compared with
the observed redshift of the galaxy to solve for the distance $r$.
However, in the vicinity of prominent overdensities, the
redshift as a function of distance can become non-monotonic, 
such that there are three different distances corresponding 
to a given redshift.  YSDH describe a distance-averaging
procedure that recognizes the existence of triple-valued zones. 
The more general approach applied here is inspired by the VELMOD 
maximum-likelihood analysis of Willick \etal (1997b). 
Along a given line of sight, 
we ask for the joint probability distribution of observing a galaxy
with redshift $cz$, flux density ${\cal F}$ and (unknown) distance $r$:
\begin{equation} 
P(cz,{\cal F},r) = P(cz|r)\times P({\cal F}|r) \times P(r)\quad.
\label{eq:joint_P} 
\end{equation}

The first term is provided by our velocity field model.
We assume that the contributions to the velocity field of 
shot noise, the nonlinearities, and the gravitational influence of material
within one smoothing length are incoherent, and thus we model 
the scatter around the assumed redshift-distance relation 
as a Gaussian, with dispersion $\sigma_v$:
\begin{equation} 
P(cz|r) = {1 \over \sqrt{2\,\pi} \sigma_v} 
\exp\left(-{
       [cz - \hat{\vx} \cdot \vv(\vx) ]^2 
        \over 2\,\sigma_v^2
           } \right)\ .
\label{eq:P(cz|r)} 
\end{equation}
For what follows, we set $\sigma_v = 150 \kms$, independent of
position (cf., Willick \etal  1997b). Like YSDH, we 
collapse clusters of galaxies to a common redshift.  
 
The second term of \equ{joint_P} 
is given by the luminosity function of galaxies, $\Phi(L)$, 
\begin{equation} 
P({\cal F}|r) = \Phi(L\!=\!4\,\pi r^2 {\cal F}) {d L /d {\cal F} } 
       \, \propto r^2 \Phi(L)\ ,
\label{eq:P(f|r)} 
\end{equation}
where the derivative is needed because the probability density is
defined in terms of ${\cal F}$, not $L$.
 
The third term in
\equ{joint_P} is given by the galaxy density distribution
along the line of sight:
\begin{equation} 
P(r) \propto n(\vx) r^2 \propto \left[1 + \delg(\vx)\right]\, r^2\ .
\label{eq:P(r)} 
\end{equation}

Finally, given the joint probability distribution, \equ{joint_P}, 
we estimate the distance of a given galaxy to be the expectation value,
\begin{equation} 
\vev{r} ={\int r P(cz,{\cal F},r)\, d r \over \int P(cz, {\cal F}, r)\, d r}\ . 
\label{eq:r-expect} 
\end{equation}

\subsection{Power-Preserving Filtering}
\label{sec:iras_ppf}

\def\delt{\delta_{\ssize T}}
\def\tdelt{{\tilde\delta}_{\ssize T}}
\def\delf{\delta_{\ssize F}}
\def\tdelf{{\tilde\delta}_{\ssize F}}
\def\delo{\delta_{\ssize O}}
\def\tdelo{{\tilde\delta}_{\ssize O}}
\def\teps{{\tilde\epsilon}}
\def\tdel{{\tilde\delta}}
\def\fw{F_{\ssize W}}
\def\fy{F_{\ssize Y}}
\def\snf{ {\vev{\delt^2} \over \vev{\epsilon^2} } }
\def\sn{ y}

  In a flux-limited sample, the mean number density of objects is a
monotonically decreasing function of distance, and therefore the shot
noise in the density field and predicted  peculiar velocity field
both increase with distance.
Several approaches to this problem have been taken in the literature,
but they all involve variable smoothing, which is not desirable 
for the purpose of comparing with the uniformly-smoothed POTENT output. 
YSDH set a top-hat smoothing length equal to the mean inter-particle spacing, 
which of course increases with distance from the origin.  Nusser \& Davis
(1995) expand the density field in spherical harmonics and radial
Bessel functions.  Given a finite number of angular modes (independent
of distance) the effective smoothing length again increases with scale.
Fisher \etal (1995) perform a similar expansion of the density field, but
in addition, they apply a Wiener filter to the expansion coefficients.  

The Wiener filter (cf., Press \etal  1992) provides the {\it minimum variance}  
reconstruction of the smoothed  density field,
given {\it a priori\/} knowledge of the underlying power spectrum and 
the noise (cf., Lahav \etal  1994; Zaroubi \etal  1995).  
Let the observed smoothed density field be 
$\delo(\vx)$, and the true underlying density field $\delt(\vx)$, such that
$\delo(\vx) = \delt(\vx) + \epsilon(\vx)$, where $\epsilon$ is the 
local contribution from shot noise.  
We wish to minimize the squared error in the density field by
filtering the observed density field.  Filtering is best done in
Fourier space, $\tdelf(\vk) =F(\vk) \tdelo$,  
where the tilde indicates Fourier Transform.
Parseval's Theorem allows us to write the squared error
of the filtered density field as
\begin{equation} 
\vev{[\delt(\vx) - \delf(\vx)]^2}
= \vev{[\tdelt(\vk) - F(\vk) \tdelo(\vk)]^2}\ .
\label{eq:variance} 
\end{equation}
Minimizing this expression with respect to the filter $F$ gives the 
{\it Wiener\/} filter, 
\begin{equation} 
\fw = { \vev{\tdelt^2} \over
\vev{\tdelt^2} + \vev{\teps^2} }\ . 
\label{eq:Wiener} 
\end{equation}
If the underlying density field and the noise are both Gaussian-distributed, 
the Wiener filter provides the most probable solution for the mean 
underlying field. 

However, the Wiener filter has a serious drawback for the present 
application. The expectation value of the square of the filtered field is:
\begin{equation} 
\vev { \fw^2\, \tdelo^2}  
= \left({\vev{\tdelt^2} 
     \over \vev{\tdelt^2} + \vev{\teps^2}}\right)^2  
     \left(\vev{\tdelt^2} + \vev{\teps^2}\right) 
= \fw \vev{\tdelt^2} \ .
\label{eq:Wiener-variance} 
\end{equation}
Since $\fw < 1$, 
the Wiener-filtered field has a variance that is always {\it smaller\/} 
than that of the true underlying field (contrast this
with the case of the observed field, whose variance is
systematically {\it greater\/} than that of the true field, by an amount
$\vev{\teps^2}$ ).  Even worse, because $\epsilon$ is in general an
increasing function of distance in a flux-limited sample,  
the variance in the Wiener-filtered
field is a decreasing function of distance.  
In fact, when the noise dominates over the signal, the most likely
value of $\delta$ tends to the mean value of the prior model, 
\ie, $\delta\rightarrow 0$.
This is unacceptable for the current application, 
because any such systematic effect in $\delg$ will translate directly 
into a bias in $\betai$ (cf., equation~[\ref{eq:gi+b}]). 

With this in mind, Yahil (1994) developed a {\it power preserving\/}
variant of the Wiener filter, defined by
\begin{equation} 
\fy = \fw^{1/2} \ .
\label{eq:PPF}
\end{equation}
A calculation analogous to that of \equ{Wiener-variance} shows that 
$\vev {\fy ^2\, \tdelo^2} = \vev{\tdelt^2}$; 
that is, the filter preserves the power in the underlying true density
field.  It is not optimal in the sense of minimum variance,
but it is straightforward to show that the ratio of the mean square 
difference between true and filtered fields for the Yahil and Wiener filters 
can be expressed in terms of the signal-to-noise ratio as
\begin{equation} 
{\vev{        (\tdelt - \fy \, \tdel)^2 } 
  \over \vev{ (\tdelt - \fw \,   \tdel)^2 } }
= 1 + \left[ \left(1+\sn\right)^{1/2} - \sn^{1/2} \right]^2 
\ , \quad \sn\equiv\snf\ . 
\label{eq:rms-ratio} 
\end{equation}
The square root of this ratio approaches unity when the signal dominates, 
$y\gg 1$.  It rises to only 1.08 when $y=1$,
and is still only 1.41 even when the noise dominates, $y\ll 1$.
That is, we pay only a mild price in density field errors when
replacing the Wiener filter by the Yahil filter,
while gaining an unbiased estimate of the variance of the density field
everywhere. 

  Applying any of these filters requires that we first define a continuous
density field.  We assign the galaxies in the sample (weighted by the 
inverse of the selection function) via cloud-in-cell to a $128^3$ 
Cartesian grid of spacing $2\hmpc$, and smooth the density field 
further with a small Gaussian window of G5 
(or G4, see \S~\ref{sec:eval_iras}).    
We then apply the Yahil filter to the Fourier Transform of the
density field, where $\vev{\delt^2}$ and $\vev{\epsilon^2}$ are 
calculated as appropriate for the smoothing window, using
as a prior the power spectrum of \iras\ galaxies as derived 
by Fisher \etal (1993).  

  Of course, the appropriate Yahil filter to apply at each point in real 
space is a function of the distance from the origin, as the noise term
$\vev{\epsilon^2}$ is an increasing function of distance.  
We thus calculate a series of Yahil-filtered density fields, 
each assuming the shot noise appropriate for one of 13 different
distances from the origin: 5, and 20 to $240\hmpc$ in steps of $20\hmpc$. 
The value of the Yahil-filtered density field at any other
point in space is then spline-interpolated from those 13 fields.

\subsection{Nonlinear Correction}
\label{sec:iras_nl}

  Because we are calculating a continuous density field, we can take
advantage of nonlinear extensions to 
\equ{gi+b}, before plugging $\divv$ into \equ{v=int_over_x} 
and solving it in Fourier space. 
Nusser \etal (1991) provided a simple functional fit to the inverse of
\equ{delc}, which does a good job of fitting nonlinear effects for 
smoothing scales of G5 and larger (see also Mancinelli \& Yahil 1995). 
The Nusser \etal expression has a mean differing from zero, which gives 
rise to an erroneous monopole term in the derived velocity field.  
Therefore, we use a generalized expression whose mean vanishes to second order:
\begin{equation} 
\divv = -\beta\, { (1 + \alpha^2 \sigma_g^2 )\, \delg + \alpha b \sigma_g^2
                  \over 1 + \alpha b^{-1} \delg } \ , 
\label{eq:Galit} 
\end{equation}
where $\sigma_g^2 \equiv \vev{\delg^2}$, and $\alpha$ is a constant
to be determined empirically.
The best fit found in CDM $N$-body simulations is $\alpha = 0.28$
(Ganon \etal 1997). 

  The iteration technique used to find self-consistent density and
velocity fields for the \iras\ sample is similar 
to that described in Strauss \etal (1992a), with the
additional application of the Yahil filter, the
nonlinear correction of \equ{Galit}, and the refined method of 
\S~\ref{sec:iras_tvz} for determining the distance of each galaxy.
We do all the calculations in the Local Group frame, 
on a $128^3$ grid centered on the observer and
oriented along Supergalactic coordinates.  The density field is
calculated within a sphere of radius $128\hmpc$, 
and is assumed to be uniform outside this sphere and inside the bounding 
cube. Periodic boundary conditions are assumed.  
We term the resulting density field of \iras\ galaxies in real space 
$\deli$, for specific assumed values of $\Omega$ and $b_I$.

\section{EVALUATING THE RECONSTRUCTION METHODS}
\label{sec:eval}

The POTENT and \iras\ reconstruction methods laid out in the previous 
sections are evaluated using mock catalogs based on $N$-body simulations. 
The systematic errors are investigated and
corrected as much as possible, and the random statistical errors
are quantified. We first describe the mock catalogs (\S~\ref{sec:eval_mock}), 
then summarize the testing of the POTENT and the \iras\ methods themselves
(\S~\ref{sec:eval_potent}, \S~\ref{sec:eval_iras}), 
and eventually describe the calibration of the comparison 
(\S~\ref{sec:method_mock}).

\subsection{The Mock Catalogs}
\label{sec:eval_mock}

The mock catalogs and the underlying $N$-body simulations
are described in detail in Kolatt \etal (1996);
we present only a brief outline here.

A special effort was made to generate simulations that mimic the actual
large-scale structure in the real universe, in order to take into account
any possible dependence of the errors on the signal.
The present-day density field, G5 smoothed, is taken to be 
that of \iras\ galaxies as reconstructed by the method described in
\S~\ref{sec:iras}, with $\bi=\Omega=1$.
The field is traced back in time to remove nonlinear effects 
by integrating the Zel'dovich-Bernoulli equation (Nusser \& Dekel 1992). 
Non-Gaussian features are removed, and structure on 
scales smaller than the smoothing length is 
added using the method of constrained realizations
(Hoffman \& Ribak 1991), with the power spectrum
of the \iras\ 1.2 Jy survey (Fisher \etal  1993) as a prior.
The resulting density field is fed as initial conditions to a PM $N$-body code
(Bertschinger \& Gelb 1991)
which then follows the nonlinear evolution under gravity, with $\Omega=1$.
The present epoch is defined by an \rms density fluctuation
of $\sigma_8=0.7$ at a top-hat smoothing of radius $8\hmpc$,
as is observed for \iras\ galaxies (Fisher \etal 1994a). 
The periodic box of side $256\hmpc$ is simulated with a $128^3$ grid
and $128^3$ particles.

Next, ``galaxies" are identified in the simulation and assigned 
the relevant physical properties.
They are then
``observed" to make mock catalogs that include all the relevant errors 
and selection effects in both the peculiar velocity and redshift survey data.  
For the mock Mark III catalogs,
each of the $N$-body particles is considered a galaxy candidate, and is
identified as elliptical or spiral, depending on the local neighborhood 
density of particles (following Dressler 1980).  
Rich clusters are identified, mimicking the cluster samples in the real data,
and the remaining particles are left as candidates for field galaxies. 
The galaxies are assigned log-linewidths $\eta$ drawn at random from
the observed $\eta$ distribution function (corresponding to the observed 
galaxy luminosity function).  A Tully-Fisher (or $D_n-\sigma$) 
relation is assumed, 
and absolute magnitudes $M$ are randomly scattered about the TF value, 
$M_{TF}(\eta)$, following a Gaussian distribution of width appropriate to the
corresponding sub-sample of the Mark III catalog.
Field galaxies are selected in the angular regions corresponding 
to each of the sub-samples, with the appropriate magnitude limits
and redshift cutoffs.
The only feature of the observational procedure that is not simulated
is the calibration of the TF relations for each sample in the Mark 
III catalog, and the matching of them in their overlap; the TF 
relations are assumed to be known perfectly {\it a priori.} 

These data are used to infer TF distances to all the
galaxies in the mock catalog.
The ``observed" redshifts are taken to be the true velocities of the
particles in the simulation.
Finally, the galaxies selected are grouped using the same code used 
for the real data, and are corrected for Malmquist bias as in
\S~\ref{sec:potent_im}, using the galaxy number density profile $n(r)$ 
as derived from a randomly selected mock \iras\ catalog 
(see \S~\ref{sec:eval_iras}).

\subsection{Errors in the POTENT Reconstruction}
\label{sec:eval_potent}

\begin{figure} [ht]
\vspace{6.7truecm}
{\includegraphics{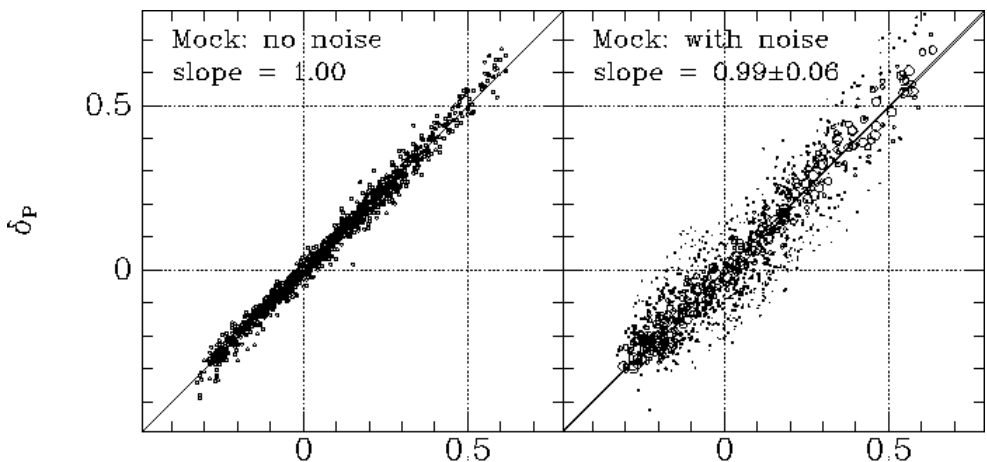}}
\caption{\capt
Systematic errors in the POTENT analysis.
The density field recovered by POTENT from the mock data
is compared with the ``true" G12 density.
The comparison is at uniform grid points within a volume of 
effective radius $40\hmpc$.
Left: The input to POTENT is the true, G12-smoothed radial velocity.
Right: The input is noisy and sparsely-sampled mock data.
Shown is $\langle \delp \rangle$,
the POTENT field averaged over 10 random realizations.
The sizes of the symbols are inversely proportional to the estimated error 
$\sp$ at that grid point.
The solid line is the average over the regression lines
in the ten realizations, of slope $0.99\pm 0.06$.
The error is the standard deviation over the realizations.  
Only half of the points (randomly selected) within the comparison 
volume are actually plotted.    
}
\label{fig:eval_potent}
\end{figure}

The left panel of Figure~\ref{fig:eval_potent} demonstrates how well
POTENT can do with ideal data of dense and uniform sampling and no
distance errors. The reconstructed density field, from input that 
consisted of the exact, G12-smoothed radial velocities, 
is compared with the true G12 density field of the simulation.
The comparison is done at grid points of spacing $5\hmpc$ inside 
our ``standard" comparison volume of effective radius $40\hmpc$ 
(see below). 
We see that no bias is introduced by the POTENT procedure itself.
The small \rms scatter of 2.5\% reflects the cumulative effects of
small deviations from potential flow,
scatter in the nonlinear approximation (equation~[\ref{eq:delc+}]),
and numerical errors  
(compare to the discussion of the nonlinear effects 
on this smoothing scale in PI93).

We execute the POTENT algorithm on each of ten noisy mock realizations of
the Mark III catalog, recovering 10 corresponding density fields.
The error in the POTENT density field at each point in space, $\sp$,
is taken to be the \rms difference over the realizations, between 
$\delp$ and $\delt$, the true G12-smoothed density field of the mass
in the simulation. 
This scatter includes both systematic and random errors. 
We evaluate the density field and the errors on 
a Cartesian grid with $5\hmpc$ spacing.
In the well-sampled regions, which extend in Mark III out to
$40\!-\!60\hmpc$, the errors are $\sp\!\approx\!0.1\!-\!0.3$, 
but they are much larger in certain regions at large distances.  
We will use $\sp$ below to exclude noisy regions in the POTENT-\iras\ 
comparison (\S~\ref{sec:method_beta}). 
In particular, our standard comparison volume is defined by $\sp\!<\!0.3$ and 
$R_4\!<\! 9.2 \hmpc$.
The resulting volume $V$ has an effective radius $R_{\rm e}=40\hmpc$,
defined by $V=(4\pi/3)R_{\rm e}^3$. 
The \rms of $\sp$ in the standard volume is 
$0.19$. 

Some part of these errors is systematic.  We can quantify this 
by comparing $\delt$ to the recovered mean density field, averaged 
over the mock catalogs at each point. 
The right panel of Figure~\ref{fig:eval_potent} makes this comparison
at the points of a uniform grid inside the standard volume.
The residuals in this scatter plot ($\la \delp \ra~ vs.~\delt$)
are the local systematic errors.
Their \rms value over the standard volume (and over the realizations)
is $0.06$. 
The corresponding \rms of the random errors ($\delp~ vs.~\la \delp \ra$) 
is $0.18$. 
The systematic and random errors add in quadrature to the 
total error ($\delp~ vs.~\delt$), whose \rms over the realizations at each
point, $\sp$, is used in the analysis below.
 
Systematic errors may also be correlated.  We quantify this by 
performing a regression of $\delp$ on $\delt$: $\delp = m\,\delt + n$,
for each realization, and then measuring the scatter around this
best-fit line: 
\begin{equation}
S^\prime \equiv {1 \over N_{\rm grid} }  \sum_i^{N_{\rm grid}}
        { \left[ {\delp}_i - (m\,{\delt}_i + n)\right]^2 \over {\sp}_i^2 } \ ,
\label{eq:Sprime}
\end{equation}
which we compute for each realization within the standard volume.
This statistic is not necessarily distributed like $\chi^2$ because
the local systematic errors may be correlated,
but it should still average to $\la S^\prime \ra =1$ 
as long as the errors of $\delp$ about $\delt$ average to zero.
The average of $S^\prime$  over the ten realizations is 
$\la S^\prime \ra = 0.94$, with a standard deviation of $0.14$
and therefore an error of $0.05$ in the mean.
The proximity of this average to unity indicates that, indeed, the
local errors roughly average to zero.

The main point of Figure~\ref{fig:eval_potent} is that the global 
systematic errors in $\delp$ as a function of $\delt$ are small.
The figure shows no systematic deviations from the $y = x$ line.
The average of the slopes $m$ of the regression lines over the ten realizations
is $0.99$ (compared to the correct answer of $1.0$) 
with a standard deviation of $0.06$.
This means that the local systematic errors show no significant correlation
with the signal, contributing no bias to the comparison with \iras. 

\begin{figure}[ht]
\vspace{6.7truecm}
{\includegraphics{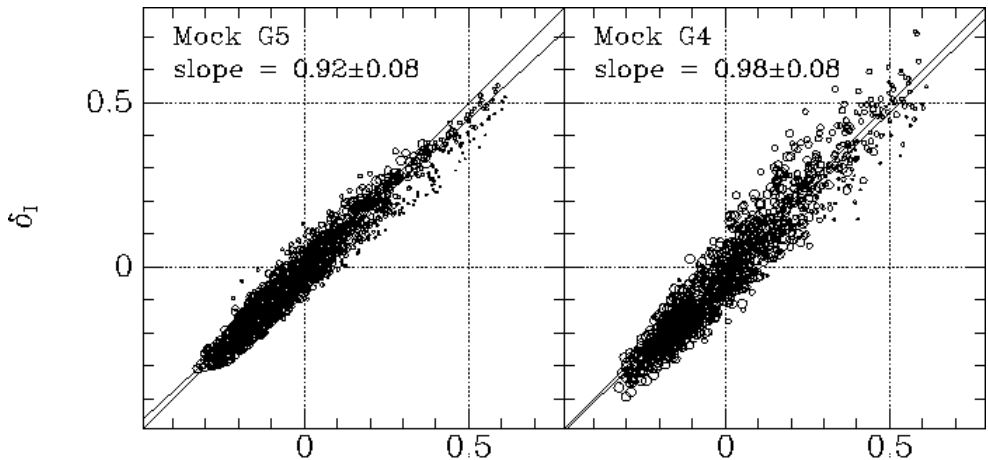}}
\caption{\capt
Systematic errors in the \iras\ analysis.
The density field reconstructed from the mock data, averaged over 9 
realizations, is compared with the ``true" G12 density.
The comparison is at uniform grid points within the standard volume of 
effective radius $40\hmpc$.
The line is the average of the regression lines over the realizations. 
The quoted slope and error are 
the average and standard deviation  
over the realizations. 
Left: with G5 smoothing in the iterative procedure, followed by G10.9
smoothing, showing a 
8\%   
($3\sigma$) bias in the slope due to over-smoothing.
Right: with G4 smoothing instead of G5 smoothing.
Only half of the points (randomly selected) within the comparison 
volume are actually plotted.   
}
\label{fig:eval_iras}
\end{figure}

\subsection{Errors in the \iras\ Reconstruction}
\label{sec:eval_iras}

Mock \iras\ redshift surveys are drawn from the simulation following 
the observed luminosity function and redshift distribution in the
\iras\ 1.2 Jy catalog. 
We allow each particle to be chosen as a galaxy with equal probability,
so $\bi = 1$ for these samples. 
Those regions of the sky not surveyed in the \iras\ sample (Strauss \etal 1990)
are excluded. 

Each of nine mock \iras\ redshift catalogs is now put through the
reconstruction code described in \S~\ref{sec:iras}, assuming
$\Omega=\bi=1$ and G5 smoothing (but see below). 
The resulting density fields are smoothed further with a Gaussian of 
$10.9\hmpc$ to yield a nominal total effective smoothing of G12, 
in order to match the smoothing used in the POTENT reconstruction. 
Finally, the density field recovered from each of the nine realizations
is compared with the true G12 density field of the simulation.

The average recovered density field is compared with the true field in
Figure~\ref{fig:eval_iras} (left panel).
The comparison is done within the standard comparison volume (defined by
\pot\ errors as before).
The average slope of the regression lines, shown in this figure, is 
$0.92\pm 0.08$,   
where the error quoted is the standard deviation over the 
nine realizations, \ie, the mean slope is about $3\sigma$ away from unity.
This bias reflects an effective over-smoothing of the \iras\ density 
field, which is partly due to the cloud-in-cell algorithm
used to place galaxies on the grid.
We find that this bias is significantly reduced by artificially 
replacing the desired G5 smoothing with a G4 smoothing.
After applying the same additional G10.9 smoothing, the average slope of the
regression becomes 
$0.98\pm 0.08$, 
which we regard as satisfactorily unbiased for the purpose of 
comparison with POTENT.
The average density, with G4+G10.9 smoothing, is compared with 
the true G12 density in the right panel of Figure~\ref{fig:eval_iras}.

The total error in the \iras\ G12 density field at each grid point,
$\si$, is taken to be the \rms value of $\deli - \delt$
over the realizations. Its 
\rms within the standard volume is
$\si=0.09$. 
The random and systematic contributions are estimated
to be $0.08$ and $0.04$ respectively.

As we did with \pot, we do regressions of each realization of $\deli$
on $\delt$, and quantify the scatter $S^\prime$, following
\equ{Sprime}.  We find $\la S^\prime\ra = 0.60$ with
a standard deviation of $\pm 0.18$. This is significantly
smaller than unity, indicating that there are systematic errors in the
determination of $\deli$ that do not average to zero.
This is not of major concern; when we later evaluate goodness of fit
of the real data using an $S^\prime$-like statistic, we do not assume
that it is distributed like $\chi^2$, but rather determine its
distribution directly from the mock realizations.

\section{METHOD OF COMPARISON} 
\label{sec:method}

\subsection{Measuring \pmb{$\betai$} }
\label{sec:method_beta}

Since the mock \iras\ and mock Mark III catalogs are both drawn from  
the same underlying density field of an $\Omega = 1$ simulation 
with $b=1$, a perfect method of comparison should yield $\beta=1$.  

As mentioned in \S~\ref{sec:eval_potent},
we limit the comparison to the regions where the recovery is reliable 
in both data sets. 
In fact, the POTENT errors are typically twice as large as the \iras\
errors, so we simply restrict ourselves to regions with small $\sp$.
In addition, in order to minimize the effects of sampling gradient bias, 
we also avoid regions of sparse sampling as indicated by a large $R_4$.

We carry out the comparison using three alternative 
sets of cuts on $\sp$ and $R_4$, defining different volumes of space
about the Local Group, in order 
to test our sensitivity to these limits and the volume surveyed. 
These three cuts are given in the left section of Table 1, which lists the
$R_4$ and $\sp$ limits, 
the number of grid points (of spacing $5\hmpc$) that satisfy this cut, the 
effective radius $R_{\rm e}$, and the
corresponding effective number of independent smoothing volumes
$N_{\rm eff}$, which we estimate by 
$N_{\rm eff}^{-1}=N_{\rm grid}^{-2}
\sum_{j=1}^{N_{\rm grid}}\sum_{i=1}^{N_{\rm grid}}\exp(-r^2_{ij}/2R_s^2)$, 
where $R_s=12\hmpc$ is the smoothing 
radius\footnote{
This $N_{\rm eff}$ is not simply the ratio of the comparison volume to the 
effective volume of the smoothing window; it also takes into account the 
shape of the comparison volume, in the sense that $N_{\rm eff}$ becomes
appropriately larger as the shape of the comparison volume deviates from a 
sphere. 
Unlike in Hudson \etal (1995), $N_{\rm eff}$ is not used in our calculations
of errors, and is provided only as a reference.}.
This range of comparison volumes represents a compromise between our
need to avoid noisy and sparsely sampled regions,
and our wish to include as large a volume as possible in order to reduce 
cosmic scatter and come closer to a fair sample.  
We have already introduced the volume of $R_{\rm e}=40\hmpc$ as our 
standard comparison volume.

Our model is that the variables $\deli$ and $\delp$ are linear
functions of the (unknown) true density field $\delt$:
\begin{equation}
\deli = \bi \delt + \epsilon_{\ssize I} \ , 
\label{eq:fac1}
\end{equation}
and
\begin{equation}
\delp = \delt + c + \epsilon_{\ssize P} \ , 
\label{eq:fac2}
\end{equation}
where $\epsilon_{\ssize I}$ and $\epsilon_{\ssize P}$ are independent
random variables with dispersions $\si$ and $\sp$, respectively.  In 
this case, we can estimate $\bi$ and $c$ by minimizing the
$\chi^2$-like quantity (Lawley \& Maxwell 1971): 
\begin{equation}
\chi ^2=\sum_i^{N_{\rm grid}} 
        { ( {\delp}_i - \bi^{-1} {\deli}_i -c )^2 
         \over {\sp}_i^2 + \bi^{-2} {\si}_i^2 } \ ,
\label{eq:chi2}
\end{equation}
with respect to $\bi$ and $c$, taking into account the errors in both fields.
The allowed offset $c$ reflects the zero-point freedom in the TF data 
and the uncertainty in the mean density in the \iras\ field.  
Remember that $\delp$ already includes the factor $f^{-1}(\Omega)$ 
(equation~\ref{eq:delc+}), which is why $\betai$ does not appear in 
this equation. 

We sample the density fields every $5\hmpc$ while our smoothing length is 
$12\hmpc$; thus the fields are greatly oversampled,  
$N_{\rm grid} \gg N_{\rm eff}$. 
It is meaningful to define a statistic, 
\begin{equation}
S = \chi^2 / N_{\rm grid} 
\label{eq:S}
\end{equation}
for estimating goodness of fit, as long as we
remember that its probability distribution function is not that of a reduced
$\chi^2$; the expectation value may be $\langle S\rangle=1$, but the dispersion
reflects $N_{\rm eff}$ and not $N_{\rm grid}$. 
We thus do not use it to determine the statistical error in $\bi^{-1}$
(as was done for example in Hudson \etal 1995). 
We instead determine the distribution of this statistic from the 
mock catalogs that satisfy our null hypotheses  
of GI and linear biasing (\S~\ref{sec:method_mock}), and then compare 
the value of $S$ obtained from the real data 
with this distribution (\S~\ref{sec:results_gof}).

\subsection{Testing the Comparison with Mock Catalogs} 
\label{sec:method_mock}

We carry out the comparison using 90 pairs of mock catalogs,
pairing the 10 mock POTENT with the 9 mock \iras\ realizations. 
For each of the three comparison volumes, 
we calculate the mean and standard deviation over the pairs 
of the quantities $S$ and $\bi^{-1}$, and list them in the middle
section of Table 1. 

\bigskip
\cl{
\begin{tabular}{cccccccccccccc} 
\multicolumn{11}{l}{ {\bf Table 1:} } \\
\hline\hline
\multicolumn{5}{c}{Comparison Volume} &&
 \multicolumn{4}{c}{Mock Data} &&
 \multicolumn{2}{c}{Real data} \\
$ \sp< $ & $R_4{}^1<$ & $N_{\rm grid}$ & $N_{\rm eff}$ & $R_{\rm e}{}^1$ &&
 $\langle S \rangle$ & $\sigma_S$ & $\langle b^{-1} \rangle$ & $\sigma_{b^{-1}}$ &&
 $S$ & $\bi^{-1}$
\\
\hline
0.20 & 8.0 & 1005 & 10 & 31 && 0.80 & 0.14 & 1.04 & 0.13 && 1.03 & 0.83\\
0.30 & 9.2 & 2081 & 18 & 40 && 0.89 & 0.16 & 1.03 & 0.12 && 1.06 & 0.89\\
0.40 & 10.0 & 3342 & 26 & 46 && 0.93 & 0.15 & 1.02 & 0.11 && 1.16 & 0.93\\
\hline\hline
$^1$$\hmpc$ 
\end{tabular}
}
\bigskip

The most important conclusion from the comparison of the mock data 
is that our method provides an almost bias-free estimate of $\bi^{-1}$.  
This is consistent with our previous tests showing that
the POTENT and \iras\ reconstructions themselves are both hardly biased
(\S~\ref{sec:eval_potent}, \ref{sec:eval_iras}).
When we combine the two, and fit for $b^{-1}$ using \equ{chi2}, 
we find that the values of $\langle b^{-1}\rangle$ 
deviate by only $2\!-\!4\%$ from the correct value of unity. 

How significant is this small bias? 
The 90 pairs of realizations are not all independent
because the \iras\ errors are smaller than the POTENT errors. 
We crudely estimate the effective number of independent pairs 
to be $\sim 20$, so the error in $\langle b^{-1} \rangle$ is roughly 
$\sigma_{b^{-1}}/\sqrt{20} \sim 0.03$,    
and thus the apparent 3\% bias is not significant. 

Another important result is that $b^{-1}$ is fairly robust
to the comparison volume in the simulations.
A factor of 3 increase in the comparison volume leads to only 
a 2\% change in the value of $\langle b^{-1}\rangle$.
One conclusion is that cosmic scatter in the simulations does not 
play a very important role in the determination of $b^{-1}$ by this method. 
Our formal error in $b^{-1}$, 
the derived scatter in $b^{-1}$ between realizations,  
does not include the cosmic scatter, and thus 
underestimates the true error, but not by a large amount.  
This cosmic scatter will be estimated again similarly in the real universe 
in \S~\ref{sec:results} by comparing results from different volumes.  

The linear fit of equation~(\ref{eq:chi2}) is affected by the
systematic errors between $\deli$ and $\delt$, and between $\delp$ and
$\delt$, that we estimated using equation~(\ref{eq:Sprime}).  Given that $\la
S^\prime \ra$ was substantially less than unity for the \iras\ reconstruction,
we expect $\si$ to be somewhat of an
overestimate, and therefore, the $S$ of the $\delp$-$\deli$ comparison
may be somewhat less than unity.  As Table~1 shows, this is indeed 
what we find for the mock catalogs. 

\begin{figure} [ht]
\vspace{6.7truecm}
{\includegraphics{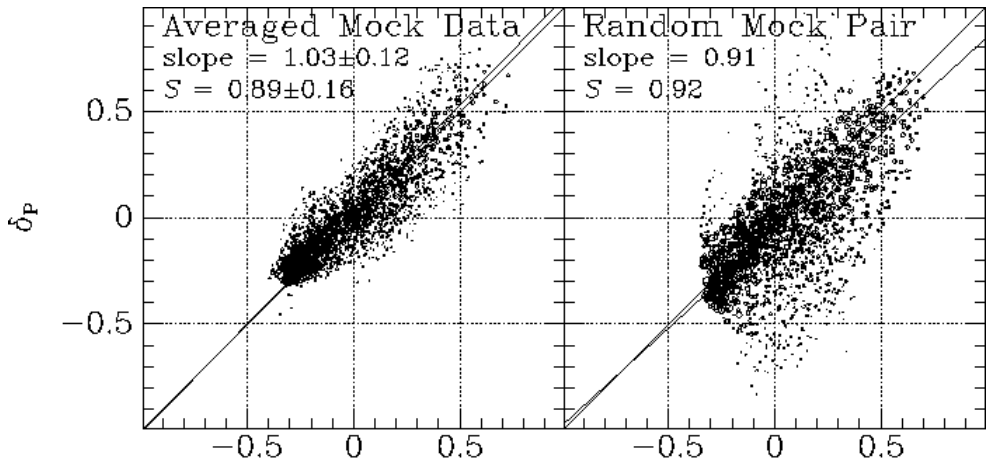}}
\caption{\capt
POTENT versus IRAS density fields as reconstructed from the mock catalogs
within the standard volume.
The symbol area is inversely proportional to the error $\se$ at each
grid point.
Left: The average mock POTENT field versus the average mock \iras\ field.
The line slope and the statistic $S$ quoted are the averages over 90 pairs.
Right: The density fields in one random pair of realizations.
}
\label{fig:pi_mock}
\end{figure}

\begin{figure}[p!]
\vskip -1truecm
\centerline{\epsfxsize=7.5 in \epsfbox{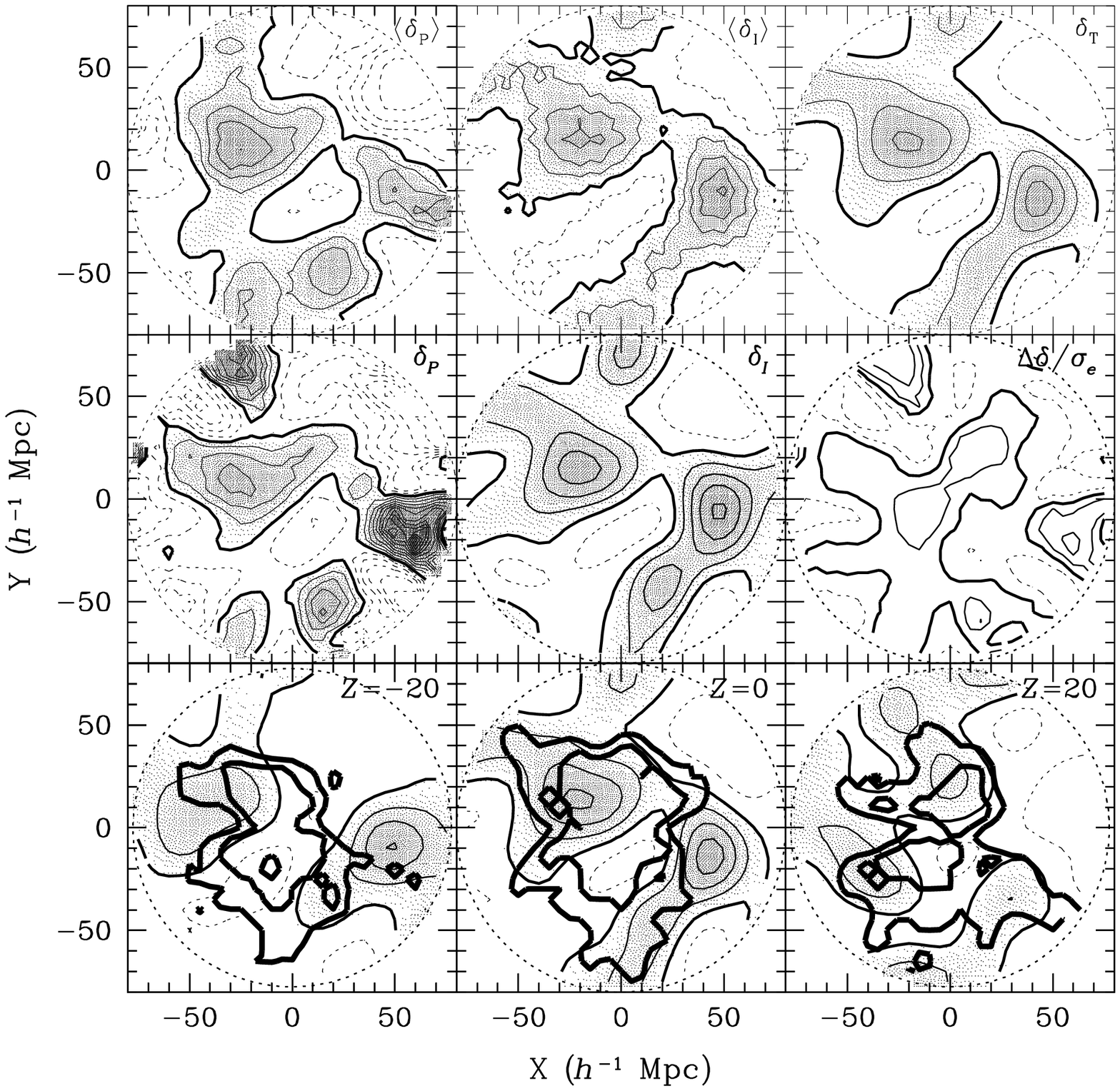}}
\vskip -0.4truecm
\caption{\capt
Density fluctuation fields of POTENT mass versus \iras\ galaxies
from the mock catalogs.
The smoothing is G12. Contour spacing is 0.2 in $\delta$,
the heavier contour is $\delta=0$, solid contours
refer to $\delta>0$ and dashed contours to $\delta<0$.
The density is also indicated by shading.
Top row: maps in the Supergalactic plane of the average fields
over the mock realizations, in comparison with the true G12 density field
of the simulation.
Middle row: the density fields reconstructed from one pair of realizations,
and the difference field $\Delta\delta$ in units of the error $\se$,
with contour spacing of unity.
Bottom row: the boundaries of the $R_{\rm e}=31$ and $46\hmpc$
comparison volumes in the Supergalactic plane ($Z=0$)
and in two parallel planes ($Z=\pm 20\hmpc$).
}
\label{fig:3x3_mock}
\end{figure}

As mentioned above, we will not use $S$ and $\chi^2$ statistics 
to determine the error in $\bi^{-1}$.  
Instead, we use the scatter of $b^{-1}$ over the pairs of mock catalogs  
as our estimate for the error in the value of $\bi^{-1}$ as derived from 
the real data.  We find that this error is $\pm 0.11-0.13$
(middle section of Table~1). 

For the offset density we find from the mock catalogs small values of 
$\langle c \rangle =0.04$, 0.03 and 0.02 
for the three comparison volumes from small to large,
with a scatter $\sigma_c=0.07$, 0.06 and 0.07 respectively.
The displacement is thus consistent with zero 
(although a small displacement is expected, especially at small volumes,
due to the fact that the mean density was determined separately 
in the reconstruction from each \iras\ mock catalog).

The left panel of Figure~\ref{fig:pi_mock}
shows a scatter plot of the average of the 
10 POTENT realizations versus the average of the 9 \iras\ realizations,
for the standard volume. 
The line of slope $1.03$, 
which is the average of the slopes over the individual pairs of 
realizations, is shown for reference (it is not necessarily identical to 
the best-fit line of $\langle\delp\rangle$ versus $\langle\deli\rangle$). 
This figure provides a visual impression of the correlation, in the 
spirit of Figure~\ref{fig:eval_potent} and Figure~\ref{fig:eval_iras}. 
The fact that the slope is close to unity reflects the lack of global bias, 
and the scatter is the local bias at the grid points. 

For a visualization of the actual 
scatter that one may expect to see in the real universe, 
we show in the right panel of Figure~\ref{fig:pi_mock} the comparison 
of one arbitrary pair of realizations. 
The scatter is naturally larger than in the left panel,
but the correlation is still very strong.

Figure~\ref{fig:3x3_mock} displays contour maps from the mock
catalogs;
the contouring is described in the figure caption. 
The upper and middle rows show G12 density 
fluctuation fields in the Supergalactic plane.
The upper row is the average over the realizations of \pot\ and \iras\
in comparison with the true G12 field, and the middle row shows 
one random realization of each and the difference between them,
in units of $\sigma_e$, defined in equation~(\ref{eq:sigmae}) below.  
It is interesting to compare these maps to Figure~\ref{fig:3x3} below, 
which shows the reconstructions of the real fields. 
Note that the \iras\ errors are 
smaller than the POTENT errors, so the mock \iras\ reconstruction
resembles the true one more closely than the POTENT reconstruction. 
The difference
between the \iras\ and POTENT fields can be very large,
especially at 
large distances, where we know the POTENT reconstruction is less reliable.
For most of the volume, the difference is of low statistical significance.

The bottom row of Figure~\ref{fig:3x3_mock} shows the boundaries
of the small ($R_{\rm e}=31\hmpc$) and large ($R_{\rm e}=46\hmpc$)
comparison volumes in the Supergalactic plane ($Z=0$)
and two parallel planes ($Z=\pm 20\hmpc$), superposed on the true field. 
The narrowing of the comparison volume in the Zone of Avoidance  
(roughly corresponding to $Y=0$), due to the large $R_4$, is clearly
visible.

\section{RESULTS} 
\label{sec:results}

With the method calibrated using the mock catalogs, 
we are in a position to compare the POTENT and \iras\ reconstructions
of the real data. We carry out the comparison
in the three comparison volumes of Table~1, using 
the errors $\sp$ and $\si$ as estimated from the mock catalogs 
in the $\chi^2$ expression of \equ{chi2}. 
We start with a visual comparison using maps (\S~\ref{sec:results_maps}),
proceed to a more quantitative analysis of the residuals in order to evaluate
the goodness of fit (\S~\ref{sec:results_gof}), 
and conclude with a measurement of $\betai$ (\S~\ref{sec:results_beta}).

\subsection {Visual Comparison}
\label {sec:results_maps}

\begin{figure}[th]
\vspace{5truecm}
{\includegraphics{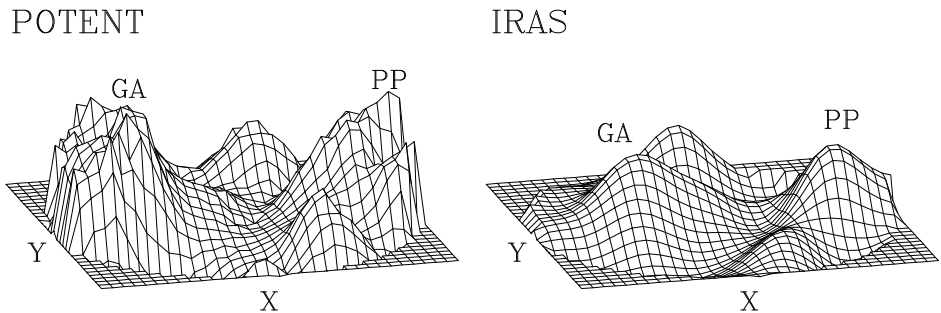}}
\caption{\capt
POTENT mass ($\Omega=1$) versus \iras\ galaxy density fields 
in the Supergalactic plane, both smoothed G12. 
The height of the surface plot is proportional to $\delta$.
The LG is at the center, GA on the left, PP on the right,
and the Sculptor void in between.  The region shown is 160\hmpc\ on a side. 
}
\label{fig:potiras_surf}
\end{figure}

Figure~\ref{fig:potiras_surf} shows the G12 density fields $\delp\,(\Omega=1)$
and $\deli$ in the Supergalactic plane. The general correlation is evident ---
the Great Attractor (GA), Perseus-Pisces (PP), and the void in between,
all show up both as dynamical entities and as structures in the galaxy
distribution.

\begin{figure}[p!]
\vskip -1truecm
\centerline{\epsfxsize=7.5 in \epsfbox{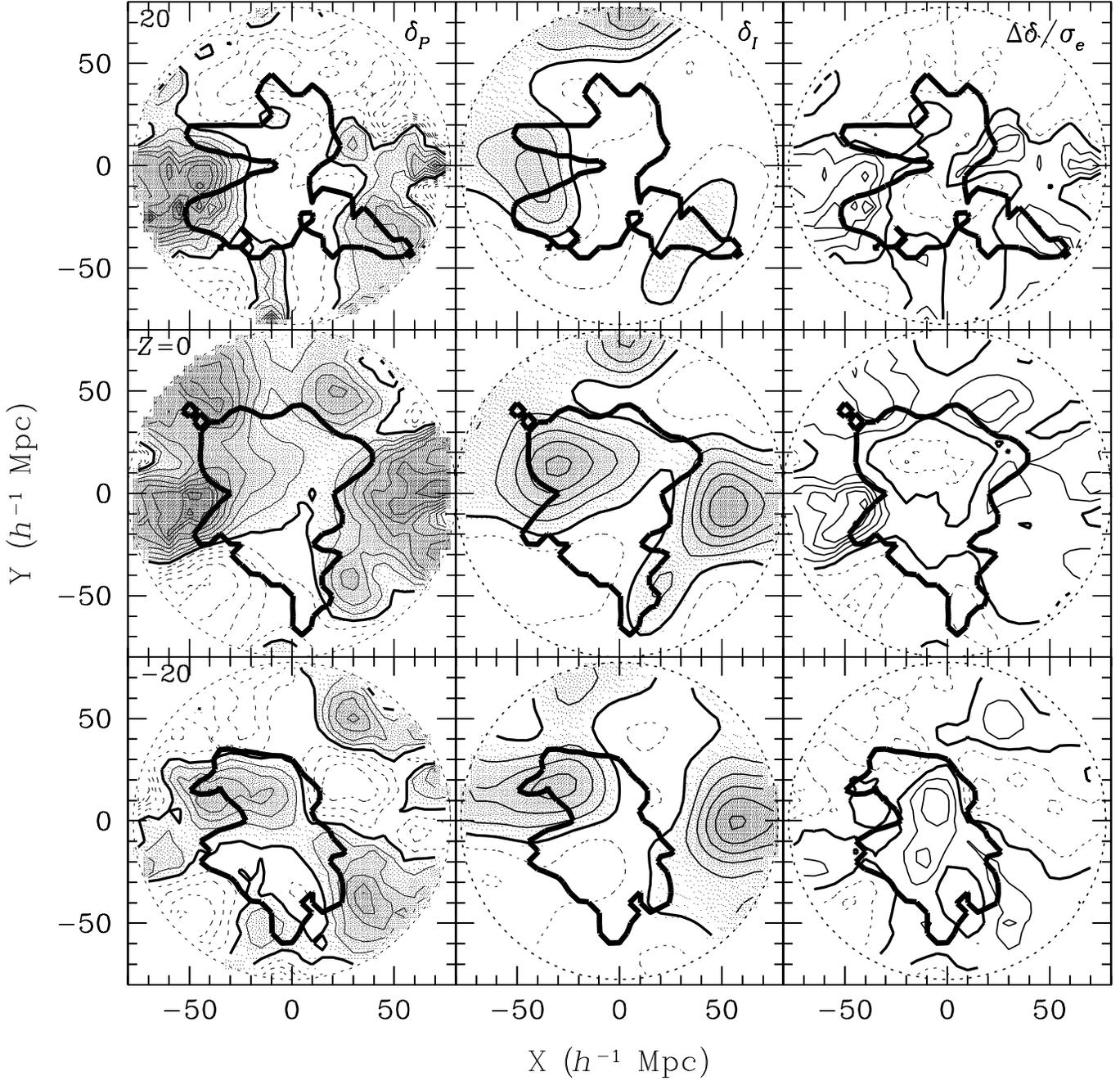}}
\vskip -0.4truecm
\caption{\capt
Density fluctuation fields of POTENT mass (left-hand column, $\Omega=1$) 
versus \iras\ galaxies (middle column), both smoothed G12.  Each row
refers to a different slice: $Z = +20 \hmpc$ (top row), $Z = 0$
(the Supergalactic plane, middle row), and $Z = -20 \hmpc$ (bottom
row). 
Contour spacing is 0.2, the heavier contour is $\delta=0$, solid contours 
mark $\delta>0$ and dashed contours $\delta<0$.
The density is also indicated by shading.
Also drawn is the difference field in units of the error, where the
contour spacing is unity
(right-hand column).
The maps are drawn out to a radius of $80\hmpc$, and
the very thick contour marks the boundaries of the $R_{\rm e}=40\hmpc$
comparison volume.  
}
\label{fig:3x3}
\end{figure}

More quantitative visual comparisons of the two density fields 
are provided in Figure~\ref{fig:3x3}, which shows in three rows
three different planes in Supergalactic coordinates:
the Supergalactic plane $Z=0$, and the planes $Z=-20\hmpc$ and
$Z=+20\hmpc$. 
The first and second columns show the G12 density fields of POTENT 
($\Omega=1$) and \iras, and the third column is the
difference between the two, 
$\Delta \delta \equiv \delp - ( \bi^{-1} \deli + c)$, 
divided by the effective error 
\begin{equation} 
\sigma_e \equiv (\sp{}^2 + \bi^{-2} \si{}^2)^{1/2}
\label{eq:sigmae}
\end{equation}
where $\bi$ and $c$ are the best-fit values
for the standard volume (\S~\ref{sec:results_beta}, Table 1).  
The contouring is described in the figure caption. 

Although the \iras\ map is noticeably less noisy than the POTENT map,
they both reveal the same main features.  The prominent density peak
near $(X,Y,Z)=(-30, 15, 0)$ \hmpc\ in the \iras\ field is centered
on the Hydra-Centaurus supercluster, which we associate with the Great
Attractor. It is elongated along the $Y$ direction because, under
G12 smoothing, the Virgo cluster [located roughly at $(15, 0, 0)$] is
an extension of the GA.  The GA is clearly seen in the POTENT field
roughly in the same region of space, except that
it peaks at a  different location $(-45, 0, 0)$ than the \iras\ peak. 
The GA is also clearly visible in the $Z=-20 \hmpc$ plane, 
where the peak appears in a similar location $(-30, 15)$ in both panels.
There is an apparently statistically significant difference
between the two, centered at ($-45$, 0, 0), but this lies in the Zone
of Avoidance, where the sampling is very sparse and
the sampling gradient bias has the strongest effect.
This is why the comparison volume does not extend to this region.
Note in particular that very little of the
volume is discrepant by more than $2\,\sigma$ (cf.,
Figure~\ref{fig:pi_distrib} below). 
In the $Z = +20 \hmpc$ plane, the GA lies outside the comparison volume.  

The extended density enhancement on the right
of the $Z=0$ maps is the Perseus-Pisces supercluster, which has
two peaks, both outside the comparison volume in either map. The PP
and GA superclusters are separated by the large Sculptor void, which
is similar in the two maps, both at $Z = 0$ and $Z = -20 \hmpc$.
Above the Supergalactic plane ($Z = +20 \hmpc$), the density field
is dominated by a large void, prominent in both POTENT and \iras. 
Another
density peak, far outside our comparison volume, is the Coma
supercluster, centered at $(5, 75, 0)$ in the \iras\ map,
and at (20, 50, 0) in the POTENT reconstruction. This 
difference need not worry us as the POTENT reconstruction is
known to be extremely noisy there.

\subsection {Goodness of Fit}
\label {sec:results_gof}

Only after we convince ourselves that the data are consistent with 
the assumed model of GI and linear biasing
can we estimate the parameter $\bi^{-1}$ and interpret it properly.
We therefore make an effort to assess goodness of fit by more 
quantitative investigations of the residuals $\Delta\delta$ 
between the POTENT and \iras\ fields. We do it in three ways;
by checking the Gaussianity of the residuals, 
by evaluating the mean amplitude of the 
residuals, and by searching for possible spatial correlations between them.

\begin{figure}[t!]
\vspace{8truecm}
{\includegraphics{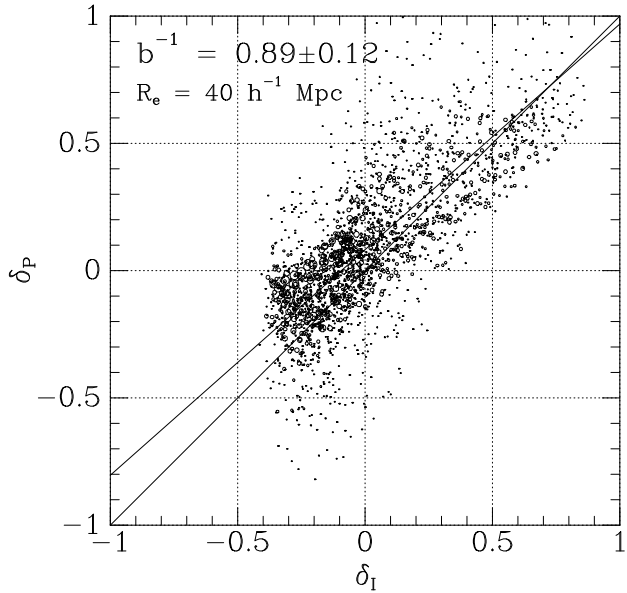}}
\caption{\capt
POTENT versus \iras\ G12 density fields from the real data,
at grid points within the standard comparison volume.
The symbol size is inversely proportional to the effective error.
The solid line is the best-fit minimizing \equ{chi2}.
}
\label{fig:scatter}
\end{figure}

A visual impression can be gained from
Figure~\ref{fig:scatter}, which shows a scatter plot of $\delp$
versus $\deli$ at the grid points within the standard volume.  
Recall that the $\sim 2000$ points are highly correlated because of the
G12 smoothing, with only $N_{\rm eff}\sim 18$ independent sub-volumes.
This gives rise to coherent clouds of points in the scatter diagram. 
Notice the qualitative similarity between this figure and the 
equivalent one for a single, arbitrary, pair of mock realizations,
Figure~\ref{fig:pi_mock}. 

\begin{figure}[t!]
\centerline{\epsfxsize=4.0 in \epsfbox{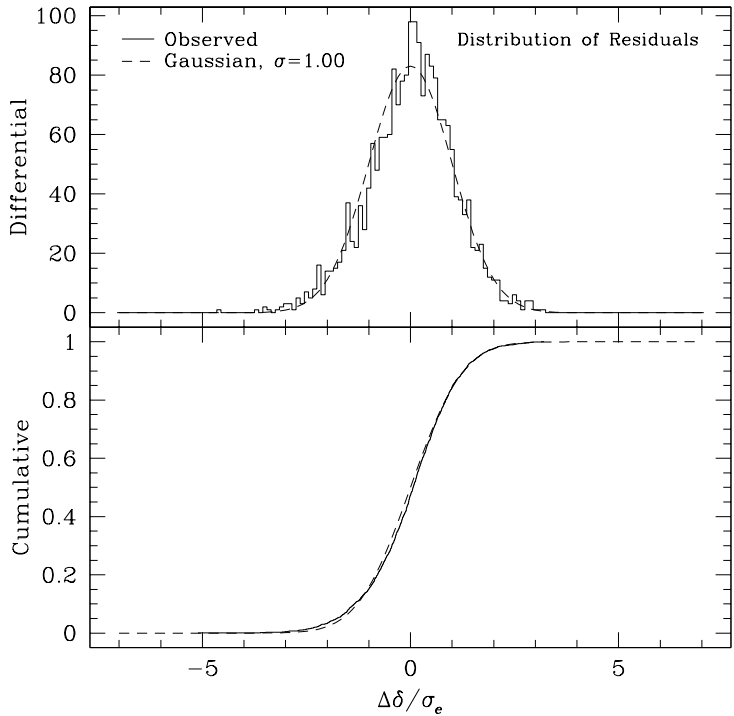}}
\caption{\capt
The probability distribution function 
of the normalized residuals, $\Delta \delta /\se$ (solid), 
in comparison with the expected Gaussian of $\sigma = 1.00$ (long-dash). 
Top: differential. Bottom: cumulative.   
}
\label{fig:pi_distrib}
\end{figure}

If we have estimated our errors accurately,
and if they are Gaussian distributed, then we expect the scatter
$\Delta \delta /\se$ 
to be Gaussian distributed with a mean of zero and a variance of unity.  
The observed differential and cumulative  
distribution function of this quantity are shown
in Figure~\ref{fig:pi_distrib} in comparison with the expected Gaussian.
The Gaussian is a reasonable match to the observed distribution;
the observed standard deviation of 1.03 (\ie, the square root of the
reduced $\chi^2$ statistic $S$, Table~1, right section) 
is only slightly different from unity.
Note in particular the reasonable fit in the tails of the distribution.
Our conclusion is that the distribution of residuals is fairly consistent 
with a Gaussian.  

\begin{figure}[t!]
\centerline{\epsfxsize=4.0 in \epsfbox{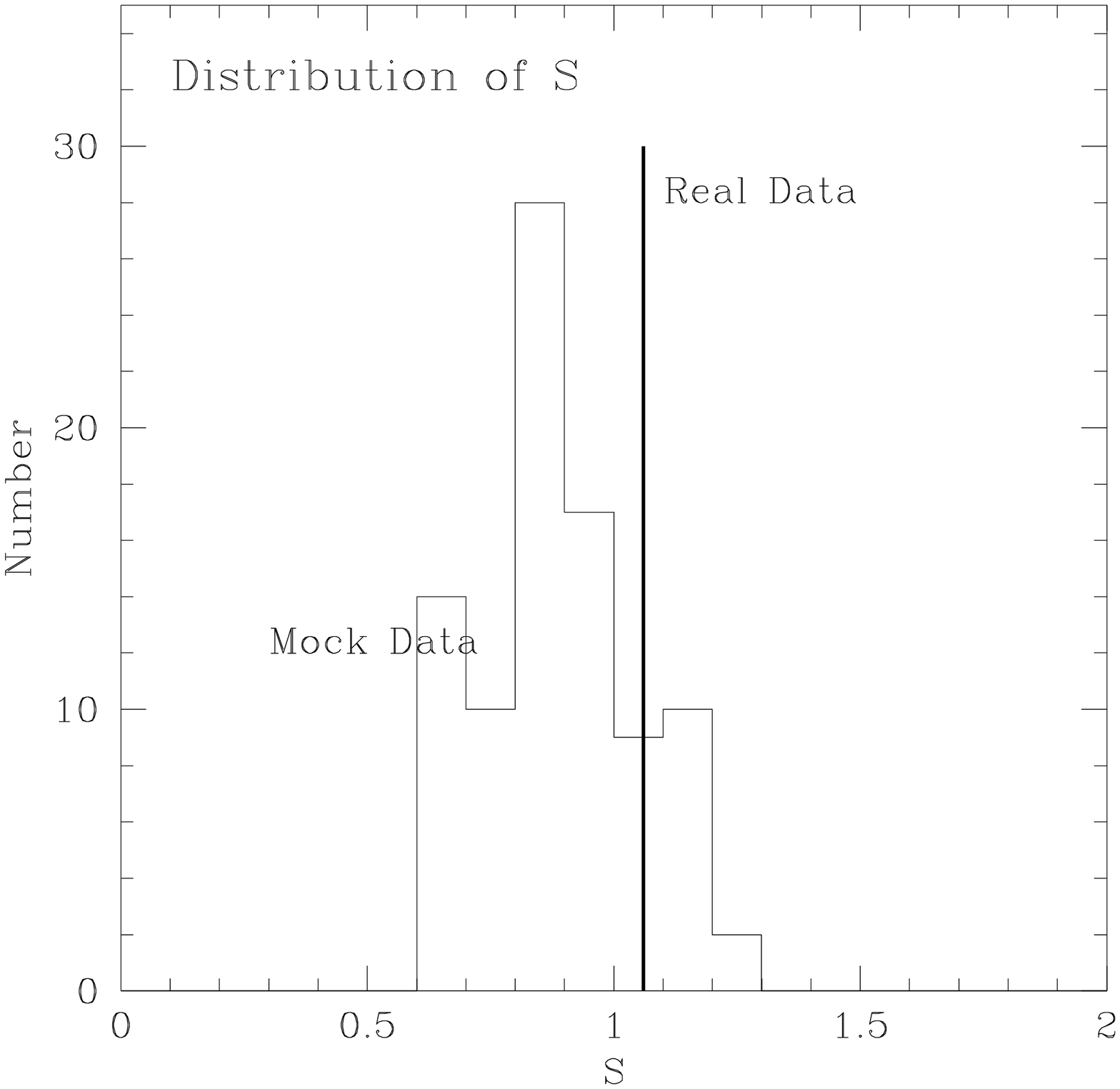}}
\caption{\capt
The observed value of $S$ within the standard volume (thick line)
in comparison with the model distribution of $S$ over the
pairs of mock realizations of the data (histogram), within the same volume.
}
\label{fig:S_data_vs_model}
\end{figure}

Figure~\ref{fig:S_data_vs_model} shows the observed value of $S$ 
within the standard volume 
in comparison with the distribution of $S$ over the 90
pairs of mock realizations of the data, representing the null
hypothesis of GI and linear biasing.
Note that this distribution has no extended tails, and would be 
well-fit by a Gaussian.  The observed value of 
$S$ falls 1$\sigma$ from the center of the distribution of $S$
in the mocks (Table 1), which is a confirmation for the goodness of fit.
For the smaller and larger comparison volumes, the observed value
of $S$ is at $1.6\sigma$ and $1.5\sigma$ respectively, indicating 
a weaker, but still acceptable goodness of fit.

\begin{figure}[t!]
\vspace{10truecm}
{\includegraphics{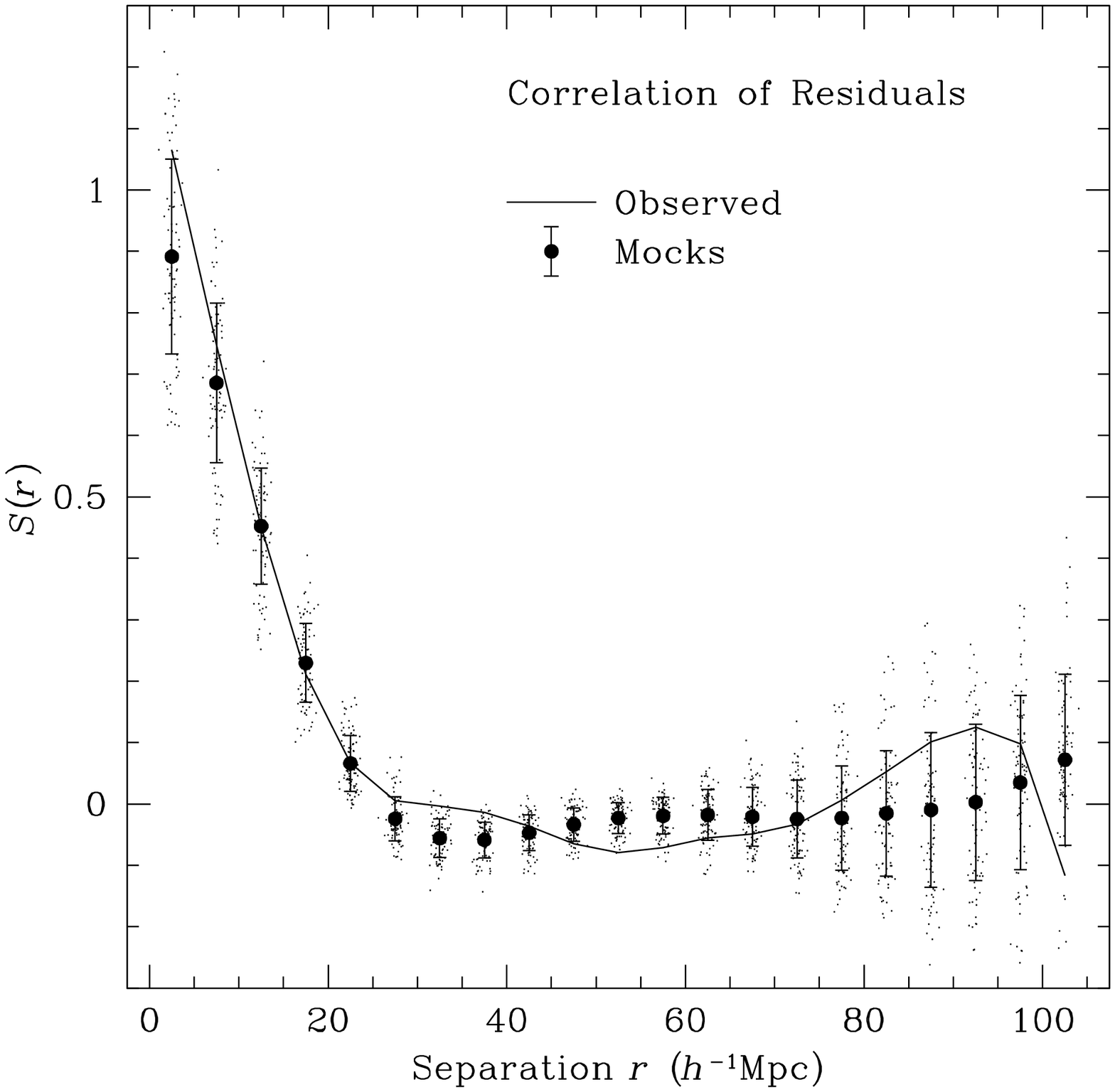}}
\caption{\capt
The correlation function $S(r)$ for the real data                  
(solid), and for the 90 pairs of mock catalogs (dots). The 
large symbols with error bars are the mean and standard deviation 
of the results from the mock catalogs at each bin of separation.     
}
\label{fig:resid-correlation}
\end{figure}

It is also important to rule out possible spatial correlations 
between the residuals, which could be 
indicative of large-scale systematic errors in the POTENT or 
\iras\ density fields.  We define a correlation function of residuals by
\begin{equation}
S(r) = \left\langle {\Delta\delta (\vx)\, \Delta\delta (\vx +\vr)
                      \over \se(\vx)\, \se(\vx +\vr)}
       \right\rangle _{\vx} \ ,
\label{eq:resid-correlation}
\end{equation}
where the average is over pairs of grid points in the comparison
volume, with pair separation in the range $(r, r+\Delta r)$, 
for $\Delta r=5\hmpc$.
Figure~\ref{fig:resid-correlation} shows this statistic for the real data
in comparison with the 90 pairs of mock catalogs.
The correlation function at zero separation coincides 
with the quantity $S$ of \equ{S} by definition,
and it is a factor of two smaller at $r \sim\! 12\hmpc$, roughly coinciding 
with the smoothing length. No statistically
significant deviation is seen between the residual correlations of
the real data and the simulations, at any separation. Thus, there is no
evidence for residual correlations in the real data beyond those
induced by the finite smoothing. 

Finally, the values for the offset between the fields are found to be
$c=0.08\pm 0.07$, $0.08\pm 0.06$ and $0.09\pm 0.07$ 
for the three comparison volumes respectively,
where the errors are the standard deviations from the mock catalogs. 
A priori, one should not be surprised to find a small offset between the 
POTENT and \iras\ density fields, because their mean densities were 
determined in somewhat different volumes and in different ways. 
These offsets are consistent with the mean and scatter found in the 
mock catalogs (\S~\ref{sec:method_mock}), 
indicating that the small offsets are not significant.

Based on the above tests, we conclude that
the \iras\ and POTENT density
field data, at G12 smoothing and within the comparison volumes, 
are {\it consistent} with the hypotheses of a GI relation
between the density and velocity fluctuation fields, and a linear
biasing relation between the density fluctuation fields
of galaxies and mass.

\subsection {The Value of \pmb{$\betai$} }
\label {sec:results_beta}

Having established a reasonable goodness-of-fit, we are now in a position to 
derive the value of $\bi^{-1}$. 
As we have assumed $\Omega = 1$ in the determination of $\delp$, this 
immediately gives us $\betai$. The results are given in Table~1.
In the standard volume we obtain $\betai=0.89\pm 0.12$.    

Had we been in the linear regime and using purely linear analysis, 
the simple relation $\divv = -\betai \deli$ would have been strictly valid, 
and the slope of the best-fit line quoted would have been a direct 
measure of $\betai$.
Since we have included mildly nonlinear corrections both in the
POTENT and the \iras\ analyses, 
we may expect the actual relation between $\delp$ and $\deli$
to deviate slightly from the linear relation.
What we have done, in fact, was to assume $\Omega=1$ in the POTENT
reconstruction (and, less importantly, $\Omega=\bi=1$ in the \iras\
conversion from redshift to real space), 
and then determine $\bi$ [and thus $\betai=f(\Omega)/\bi$]
from the best fit line of $\delp$ versus $\deli$.
The fact that we ended up with a value of $\betai$ that is consistent 
at the $\sim 1\sigma$ level with the assumed value of unity 
indicates that this is a self-consistent solution
for $\Omega$ and $\bi$ near unity.

In order to validate our result for the more general case of $\betai \sim 1$ 
where $\Omega$ and $\bi$ may differ from unity, 
one should redo the $N$-body simulation of the mock catalogs 
for values of $\Omega \ne 1$. 
In the present paper, we estimate the errors for $\Omega < 1$ by scaling
the errors determined from the $\Omega=1$ simulation. 
We first redo the comparison analysis in the standard volume
with assumed values of $\Omega=0.3$ and $\bi=0.5$, which also 
correspond to $\betai=1$. 
This makes a difference mostly in the POTENT reconstruction.
We crudely scale the errors $\sp$ in proportion to $f(\Omega)^{-1}$.
For this case we obtain  $\betai=1.00\pm 0.18$   
from our fit to the real data in our standard volume. 
The quoted error is now only a rough estimate based on the increase in $\se$. 
This result is consistent with the value $\betai=0.89\pm 0.12$,    
obtained in our standard case of $\Omega=\bi=1$. 
This confirms our assessment that we are indeed measuring $\betai$, 
and are not very  sensitive to the actual values of $\Omega$ and $\bi$
that give rise to that value of $\betai$.

Finally, we try two cases where the assumed $\betai$ is $\sim 0.5$ rather
than unity.  When we start with $\Omega=1$ and $\bi=2$, we obtain 
$\betai=0.93\pm 0.12$.   
When we start with $\Omega=0.3$ and $\bi=1$, we obtain 
$\betai=0.90\pm 0.18$       
where the error is roughly scaled as before. 
In both cases, the resultant $\betai$ are significantly larger than the
``assumed" value of $\betai=0.5$.
This test demonstrates that the result is not determined by the
assumed input and the method is discriminatory; we can rule out low values
of $\betai\leq 0.5$. Note, however, that the rejection of the case $\bi=2$
is stronger than the rejection of the case $\Omega=0.3$, because the
\pot\ errors become larger when $\Omega$ is lower.

The above tests strengthen our confidence in the generality
of our result from the standard case, of 
$\betai \simeq 0.89$.         
However, the error estimate of 
$\pm 0.12$    
is strictly valid only
near $\Omega \sim 1$. If $\Omega$ is lower, the error
is expected to be larger.  
The calibration of the method and the error estimate
can be fine-tuned further using full POTENT and \iras\ reconstructions
from mock catalogs drawn from simulations with different values of 
$\Omega$ and $\bi$, which is work in progress.  
However, since we have no reason to suspect any significant change in 
the result, we defer this further testing to a subsequent paper.

\section{DISCUSSION AND CONCLUSIONS}
\label{sec:conc}

We have compared the density of mass and light as reconstructed 
from the Mark III catalog of peculiar velocities and the \iras\ 1.2 Jy 
redshift survey, with Gaussian smoothing of $12\hmpc$ 
and within volumes of effective radii $31-46\hmpc$.
Our two main conclusions are:
\begin{itemize}
\item{}
The data are consistent with gravitational instability theory in the
mildly nonlinear regime and a linear biasing relation between
galaxies and mass on these large scales.
\item{}
The value of the corresponding $\beta$ parameter at this smoothing scale is
$\betai=0.89 \pm 0.12$.     
The relative robustness of this value to changes 
in the comparison volume suggests that the cosmic scatter
is no greater than $\pm 0.1$. 
\end{itemize}
This result is consistent with the simplest cosmological model of an
Einstein -- de Sitter universe ($\Omega=1$, $\Lambda=0$) with unbiased 
\iras\ galaxies, and it also permits somewhat lower values of $\Omega$ and 
$\bi$.  However, values as low as $\Omega\sim 0.3$ would require large-scale
anti-biasing of $\bi<0.75$ (with 95\% confidence) --- a phenomenon 
that is not easily reproduced in theoretical simulations (e.g., Cen \&
Ostriker 1992; Evrard, Summers, \& Davis 1994). 

  Although our analysis uses an extension of the Zel'dovich
approximation, we have not tried to directly measure nonlinear
effects in these data to break the degeneracy between $\Omega$ and
$\bi$: the effects are weak (PI93), and are of the same order as 
possible nonlinearities in the biasing relation.  An interesting area
for future work is to see if we can put an {\it upper\/} limit on the
degree of nonlinearity; if we can, this puts a lower limit on $\bi$,
and therefore a lower limit on $\Omega$. 

As explained in \S~\ref{sec:intro},
the current analysis is a significantly improved version of the PI93
analysis done with the earlier Mark II and \iras\ 1.936 Jy data,
and of the Hudson \etal (1995) comparison to optical galaxies.
The result of PI93 was $\betai=1.28^{+0.75}_{-0.59}$ at 95\% 
confidence. The new result is lower, but only by about $1.3\sigma$. 
The current analysis is superior in many respects. 
The improvements in the data include denser sampling of a larger volume,
a careful procedure of selection, calibration and merging of the TF data
sets, and a better correction for Malmquist bias.
The methods of reconstruction were considerably improved,
using new techniques and with extensive testing and error
analysis using realistic mock catalogs drawn from simulations. 
The optimization of the comparison method using the mock catalogs led
to an unbiased estimate of $\beta$ with well-defined statistical errors.

The quantity $\beta$ has been measured using a variety of techniques
from observational data sets (as discussed in the references in
\S~\ref{sec:intro}); we here discuss other estimates of $\betai$ from
the Mark III and \iras\ 1.2 Jy data. 

Estimates of $\betai$ from {\it redshift distortions\/} in redshift
surveys are almost all within 2$\sigma$ of our current result 
(Peacock \& Dodds 1994,  $\betai = 1.0\pm0.2$;
Fisher \etal 1994b, $\betai = 0.45^{+0.3}_{-0.2}$;
Fisher, Sharf, \& Lahav 1994c, $\betai = 1.0 \pm 0.3$;
Cole, Fisher, \& Weinberg 1995, $\betai = 0.5 \pm 0.15$;
Hamilton 1995, $\betai = 0.7 \pm 0.2$;
Heavens \& Taylor 1995, $\betai = 1.1 \pm 0.3$;
Fisher \& Nusser 1996, $\betai = 0.6 \pm 0.2$).
The scatter between these results is mostly due to differences in method of 
analysis, but 
cosmic scatter, nonlinear effects, and complications in the biasing
scheme probably also play a role. 

Our current analysis is a comparison at the density level 
(a $\delta-\delta$ comparison).
Alternative analyses that performed the comparison at the {\it velocity\/} 
level ($v-v$ comparisons) typically yielded somewhat lower values 
for $\betai$, ranging from $0.5$ to $0.9$:
(Strauss 1989, $\betai \simeq 0.8$;
Kaiser \etal 1991, $\betai = 0.86^{+0.2}_{-0.15}$;
Roth 1994, $\betai = 0.6\pm0.2$;
Nusser \& Davis 1994, $\betai = 0.6\pm 0.2$;
Davis \etal 1997, $\betai = 0.6\pm 0.2$;
Willick \etal 1997b, $\betai = 0.49 \pm 0.07$).
We focus here on the two most recent 
and sophisticated analyses of the $v-v$ type, termed ITF and VELMOD.

The ITF analysis (Davis \etal 1997) is a mode-by-mode comparison
of velocity-field models, based on expansion in spherical harmonics and
radial Bessel functions, that were fitted in redshift space to a processed 
version of the inverse TF data from the Mark III catalog and the 
\iras\ 1.2 Jy redshift survey.  The ITF analysis finds a poor 
fit between the velocity fields and the model, 
in the form of a distance-dependent dipole at large distances, 
in apparent contrast with the good fit obtained here. 
One possible explanation for this difference is an error in one of the 
large-scale velocity fields, or both, that does not propagate to their 
local derivatives, the density fields.
A differential offset in the zero point of the TF relation between
data sets of the Mark III catalog, which generates an artificial coherent
bulk motion, is one such possibility. 
An inaccurate definition of the Local Group frame in the \iras\ 
reconstruction, or missing density data beyond the sampled volume,
are other possible sources of error in the velocities. 
Another source of error is the somewhat arbitrary adjustments that had
to be made to the Mark III data sets when combining them into one system 
for the ITF analysis
(cf., the discussion in Davis \etal 1997). 
If one ignores the poor fit, the ITF analysis yields the formal estimate  
$\betai = 0.6 \pm 0.2$, 
$\sim 1 \sigma$ lower than our current result.
The local nature of the density-density comparison, and the resulting
better goodness of fit, argue that it may provide a more reliable
estimate of $\betai$.

The VELMOD analysis (Willick \etal 1997b) is a high-resolution
$v-v$ comparison at a small smoothing scale of $3\hmpc$, 
using about one quarter of the Mark III data within a volume of 
radius $< 30\hmpc$, and applying a sophisticated likelihood analysis 
in the linear regime.
The comparison revealed a good fit and an estimate of $\betai=0.49\pm0.07$.
We can see several possible reasons for the difference in the estimates
of $\betai$.

It is possible that the difference originates from 
systematic errors that were somehow overlooked despite the successful
testing using the same mock catalogs. For example, the VELMOD analysis
assumes pure linear theory, 
and may thus suffer from nonlinear effects that were not taken into account. 
In fact,
it was never fully understood how the linear analysis of VELMOD did well 
at G3 smoothing where nonlinear effects are expected to be important.
One suspicion is that the mock simulation is too smooth on small scales.

Another source for the difference may be the partial data used in the 
VELMOD comparison, where the comparison is in fact dominated by data within 
$\sim 20\hmpc$.  The difference in $\betai$ can thus partly reflect cosmic 
scatter in $\bi$ or in the local effective $\Omega$. 
Table~1 shows that $\betai$ increases by an insignificant $1\sigma$ 
from the smallest to the largest comparison volume; the VELMOD
analysis also found no statistically significant growth of $\betai$
with scale.  This allows us to put an upper limit on the cosmic
scatter of order 0.1. 

Finally, a possible explanation for the difference in $\betai$ is {\it scale
dependence} of the biasing relation 
between Gaussian smoothings of $3$ and $12\hmpc$,
which would be associated with non-linear biasing.  
Some support for such a trend is found by the SIMPOT analysis
(Nusser \& Dekel 1997), which fits a parametric model of
velocity and density fields and $\beta$ simultaneously to the peculiar
velocity and redshift data (a $v-\delta$ comparison).
This analysis yields $\betai =1.0\pm0.15$ for G12 smoothing, 
and lower values of $\betai$ for smaller smoothing scales,
in qualitative agreement with the comparison of the results 
of the current paper and of VELMOD.
Current theoretical simulations indicate possible scale dependence 
in the biasing relation between scales of one to a few megaparsecs
(\eg, Kauffman \etal 1997; cf., Mo \& White 1996), but it remains to 
be seen whether such a trend can arise on scales of $6$ to $12\hmpc$.
The biasing scheme, which could be nontrivial in several ways
(see Dekel \& Lahav 1997), is clearly a bottle-neck in the effort 
to measure the cosmological parameter $\Omega$ via $\beta$ --- 
the biasing is inevitably involved whenever galaxy-density data are used.

  The largest source of error in this analysis lies in the peculiar
velocity data, and thus improved peculiar velocity datasets, with
better control of systematic errors, denser sampling, and more
complete sky coverage, will be of great importance for this work (cf.,
the reviews of Strauss 1997a; Giovanelli 1997).
It will be interesting to make the
comparison of the POTENT maps with the optical redshift data of
Santiago \etal (1995, 1996), although a proper calculation of the
errors in the optical density field will be somewhat more difficult. 
Finally, more work lies ahead in testing the robustness of our results
to the assumed value of $\Omega$ (see the discussion in
\S~\ref{sec:results_beta}), and carrying out the analysis at
other smoothing lengths to look for nonlinear effects. 

\acknowledgments{
We acknowledge Tsafrir
Kolatt for his work on the mock catalogs, Galit Ganon for her work on
the nonlinear approximations, and Idit Zehavi for thoughtful comments on the
errors.
We thank the rest of the Mark III team, David Burstein, St\'ephane Courteau, 
Sandy Faber, and Jeffrey Willick. This paper benefited from earlier
work with Edmund Bertschinger, Marc Davis, Michael Hudson, and Adi Nusser.
This research was supported in part by the US-Israel Binational Science 
Foundation grant 95-00330, by the Israel Science Foundation grant 950/95,
by NSF Grant AST96-16901, and by a NASA Theory grant at UCSC. 
MAS gratefully acknowledges the support of an Alfred P. Sloan Foundation
Fellowship. 
}


\def\re{\reference}
\def\jeru{in {\it Formation of Structure in the Universe},
     eds.~A. Dekel \& J.P. Ostriker (Cambridge Univ. Press)\ }

\end{document}